\documentclass[twocolumn,epjc3]{svjour3}

\usepackage{amsmath,amssymb}

\usepackage{ulem}
\usepackage{url}
\usepackage{bm}
\usepackage{graphicx}
\usepackage{epsfig}
\usepackage{rotating}
\usepackage{dcolumn}
\usepackage{xcolor}
\usepackage{slashed}
\usepackage{threeparttable}
\usepackage{booktabs}
\usepackage[colorlinks=true
,urlcolor=blue
,anchorcolor=blue
,citecolor=blue
,filecolor=blue
,linkcolor=blue
,menucolor=blue
,pagecolor=blue
,linktocpage=true
,pdfproducer=medialab
]{hyperref}

\RequirePackage{mathptmx}
\RequirePackage{latexsym}
\RequirePackage[numbers,sort&compress]{natbib}

\definecolor{forestgreen}{HTML}{009B55}

\newcommand{\beq}{\begin{eqnarray}} 
\newcommand{\eeq}{\end{eqnarray}} 
\allowdisplaybreaks[1]

\journalname{Eur. Phys. J. C}

\begin{document}
\title{
\begin{minipage}{\textwidth}
\vspace*{-1.0cm}
\begin{flushright}
{\tt \normalsize CERN-TH-2022-108 \\
IFIC/23-08 \\
KA-TP-02-2023\\
P3H-23-013 \\[-0.2cm]
PSI-PR-23-6}
\end{flushright}
\vspace*{0.0cm}
\end{minipage}
{\boldmath 
Full NLO QCD predictions for Higgs-pair production in the
2-Higgs-Doublet Model}
}


\author{
  J.\,Baglio\thanksref{e1, addr1, addr2}
  \and
  F.\,Campanario\thanksref{e2, addr3}
  \and
  S.\,Glaus\thanksref{addr4, addr5}
  \and
  M.\,M\"uhlleitner\thanksref{e4, addr4}
  \and
  J.\,Ronca\thanksref{e5, addr6}
  \and
\parbox[t]{1.5cm}{M.\,Spira\thanksref{e6, addr7}}
}
\thankstext{e1}{email: \href{mailto:julien.baglio@cern.ch}{julien.baglio@cern.ch}}
\thankstext{e2}{email: \href{mailto:francisco.campanario@ific.uv.es}{francisco.campanario@ific.uv.es}}
\thankstext{e4}{email: \href{mailto:milada.muehlleitner@kit.edu}{milada.muehlleitner@kit.edu}}
\thankstext{e5}{email: \href{mailto:jonathan.ronca@uniroma3.it}{jonathan.ronca@uniroma3.it}}
\thankstext{e6}{email: \href{mailto:michael.spira@psi.ch}{michael.spira@psi.ch}}


\institute{Theoretical Physics Department, CERN, CH-1211
  Geneva 23, Switzerland\label{addr1}
  \and
  QuantumBasel, uptownBasel Infinity Corp., Schorenweg 44B, CH-4144 Arlesheim, Switzerland\label{addr2}
  \and
  Theory Division, IFIC, University of Valencia-CSIC, E-46980 Paterna,
  Valencia, Spain\label{addr3}
  \and
  Institute for Theoretical Physics, Karlsruhe Institute of Technology,
  D-76128 Karlsruhe, Germany\label{addr4}
  \and
  Institute for Nuclear Physics, Karlsruhe Institute of Technology,
  D-76344 Karlsruhe, Germany\label{addr5}
  \and
  Dipartimento di Matematica e Fisica, Universit\`a degli Studi Roma Tre, and INFN, Sezione di Roma Tre, I-00146 Rome, Italy\label{addr6}
  \and
  Laboratory for Particle Physics, Paul Scherrer Institut, CH-5232
  Villigen PSI, Switzerland\label{addr7}}

\date{\today}

\maketitle

\begin{abstract}
  After the discovery of the Higgs boson in 2012 at the CERN Large 
  Hadron Collider (LHC), the study of its properties still leaves room
  for an extended Higgs sector with more than one Higgs
  boson. 2-Higgs Doublet Models (2HDMs) are well-motivated extensions
  of the Standard Model (SM) with five physical Higgs bosons: two
  CP-even states $h$ and $H$, one CP-odd state $A$, and two charged
  states $H^\pm_{}$. In this letter, we present the calculation of the
  full next-to-leading order (NLO) QCD corrections to $hH$
  and $AA$ production at the LHC in the 2HDM at small
  values of the ratio of the vacuum expectation values, $\tan\beta$,
  including the exact top-mass dependence everywhere in the
  calculation. Using techniques applied in the NLO QCD SM Higgs pair
  production calculation, we present results for the total cross
  section as well as for the invariant Higgs-pair-mass distribution at
  the LHC. We also provide the top-quark scale and scheme
  uncertainties which are found to be sizeable.
\end{abstract}

\section{Introduction}

2-Higgs Doublet Models \cite{Lee:1973iz,Branco:2011iw} are well
motivated extensions of the SM. They belong to the simplest Higgs sector
extensions of the SM that, taking into account all relevant theoretical
and experimental constraints, are testable at the LHC. In their type II
version they contain the Higgs sector of the Minimal Supersymmetric
extension of the SM (MSSM) as a special case. Featuring five physical
Higgs bosons after electroweak symmetry breaking (EWSB), they represent
an ideal benchmark framework for the investigation of various possible
new physics effects to be expected at the LHC in multi-Higgs boson
sectors.

The neutral Higgs boson pairs of the 2HDM are dominantly produced via
the loop-induced gluon-fusion process $gg\to\phi_1\phi_2$, where
$\phi_{1/2}$ denote scalar or pseudoscalar Higgs bosons of the 2HDM.
Only for mixed scalar+pseudoscalar Higgs production the Drell--Yan-type
process $q\bar q\to Z^*\to A+h/H$ takes over the dominant role in large
regions of the parameter space \cite{Dawson:1998py}. The topic of our
paper is the calculation of the full NLO QCD corrections to scalar
Higgs-pair and pseudoscalar Higgs-pair production via gluon fusion
within the 2HDM.

In the past the NLO QCD corrections to the gluon-fusion process $gg\to
HH$ have been calculated within the SM and the MSSM in the heavy-top
limit (HTL) \cite{Dawson:1998py}. This calculation has been extended to
the NNLO QCD corrections in the HTL \cite{deFlorian:2013uza,
deFlorian:2013jea, Grigo:2014jma}. Quite recently, this level has been
extended to the N$^3$LO order in the HTL \cite{Banerjee:2018lfq,
Chen:2019lzz, Chen:2019fhs, Spira:2016zna}. On the other hand finite top
mass effects beyond the HTL have turned out to be sizeable
\cite{Borowka:2016ehy, Borowka:2016ypz, Baglio:2018lrj, Baglio:2020ini,
Baglio:2020wgt}. The inclusion of the related uncertainties due to the
scheme and scale dependence of the virtual top mass has been shown to be
mandatory, since they dominate the intrinsic theoretical uncertainties
\cite{Baglio:2018lrj, Baglio:2020ini, Baglio:2020wgt}. For BSM
scenarios, the NLO QCD corrections to all production modes involving
scalar and pseudoscalar Higgs bosons are known in the HTL
\cite{Dawson:1998py}, while partial results for the virtual corrections
to pseudoscalar Higgs-pair production are known beyond NLO QCD within
the HTL \cite{Bhattacharya:2019oun}.

The paper is organised as follows. In Section~\ref{sec:model} we
introduce the 2HDM and the benchmark point we have selected to obtain
our numerical results, then we give a short description of the details
of our calculation in Section~\ref{sec:calculation}. Our results for
$hH$ and $AA$ production are presented in Section~\ref{sec:numresults}.
The theoretical uncertainties are discussed in Section~\ref{sec:errors},
in particular the top-quark scale and scheme uncertainties in
Section~\ref{subsec:topquark}. A short conclusion is given in
Section~\ref{sec:conclusion}.

\section{The 2-Higgs Doublet Model}
\label{sec:model}
The 2HDM is obtained by extending the SM by a second Higgs doublet
with the same hypercharge. We work within the 2HDM version with a
softly broken $\mathbb{Z}_2$ symmetry under which the two Higgs
doublets $\Phi_{1,2}$ behave as $\Phi_1 \to -\Phi_1$ and $\Phi_2 \to 
\Phi_2$. In terms of the two $SU(2)_L$ Higgs doublets with hypercharge
$Y=+1$ the most general scalar potential that is invariant under the
$SU(2)_L \times U(1)_Y$ gauge symmetry and that has a softly broken
$\mathbb{Z}_2$ symmetry is given by
\beq
V &=& m_{11}^2 |\Phi_1|^2 + m_{22}^2 |\Phi_2|^2 - m_{12}^2 (\Phi_1^\dagger
\Phi_2 + h.c.) \nonumber\\
&& + \frac{\lambda_1}{2} (\Phi_1^\dagger \Phi_1)^2 +
\frac{\lambda_2}{2} (\Phi_2^\dagger \Phi_2)^2 
+ \lambda_3
(\Phi_1^\dagger \Phi_1) (\Phi_2^\dagger \Phi_2) \nonumber \\
&& + \lambda_4 
(\Phi_1^\dagger \Phi_2) (\Phi_2^\dagger \Phi_1) + \frac{\lambda_5}{2}
[(\Phi_1^\dagger \Phi_2)^2 + h.c.] \;.
\eeq
Working in the CP-conserving 2HDM, the three mass parameters,
$m_{11}$, $m_{22}$ and $m_{12}$, and the five coupling parameters
$\lambda_1$-$\lambda_5$ are real. The discrete
  $\mathbb{Z}_2$ symmetry (softly broken by the term proportional to $m_{12}^2$)
has been introduced to ensure the absence of tree-level flavour-changing
neutral currents (FCNC). Extending the $\mathbb{Z}_2$ symmetry to the
fermion sector, all families of same-charge fermions will be forced to
couple to a single doublet so that tree-level FCNCs will be eliminated
\cite{Branco:2011iw,Glashow:1976nt}. This implies four different types
of doublet couplings to the fermions that are listed in
Table~\ref{tab:model1} together with the transformation properties of
the fermions. The corresponding 2HDM types are named type~I, type~II,
lepton-specific and flipped. The resulting couplings of the fermions
normalised to the SM couplings can be found in \cite{Branco:2011iw}.
\begin{table}[b!]
 \begin{center}
 \begin{tabular}{c|ccc|ccccc}
     \toprule
Model & $u_R$ & $d_R$ & $e_R$ & $Q$ & $u_R$ & $d_R$ & $L$ & $l_R$\\
   \midrule
type I & $\Phi_2$ & $\Phi_2$ & $\Phi_2$ & $+$ & $+$ & $+$ & $+$ & $+$ \\
type II & $\Phi_2$ & $\Phi_1$ & $\Phi_1$ & $+$ & $+$ & $-$ & $+$ & $-$ \\ 
flipped & $\Phi_2$ & $\Phi_1$ & $\Phi_2$ & $+$ & $+$ & $-$ & $+$ & $+$ \\
lepton-specific & $\Phi_2$ & $\Phi_2$ & $\Phi_1$ & $+$ & $+$ & $+$ & $+$ & $-$ \\
   \bottomrule
  \end{tabular}
\caption{Classification of the Yukawa types of the $\mathbb{Z}_2$
  symmetric 2HDM. 2nd-4th columns: allowed coupling combinations of Higgs
  doublet and fermion types; last
  five columns: $\mathbb{Z}_2$ 
  assignments for the quark doublet $Q$, the up-type quark singlet
  $u_R$, the down-type quark singlet $d_R$, the lepton doublet $L$,
  and the lepton singlet $l_R$. \label{tab:model1}}
   \end{center}
 \end{table}
After EWSB, the Higgs doublets $\Phi_i$ $(i=1,2)$ can be expressed in
terms of their vacuum expectation values (VEV) $v_i$, the charged complex fields 
$\phi_i^+$, and the real neutral CP-even and CP-odd fields $\rho_i$ and
$\eta_i$, respectively, as  
\beq
\Phi_1 = \left(
\begin{array}{c}
\phi_1^+ \\
\frac{\rho_1 + i \eta_1 + v_1}{\sqrt{2}}
\end{array}
\right) \qquad \mbox{and} \qquad
\Phi_2 = \left(
\begin{array}{c}
\phi_2^+ \\
\frac{\rho_2 + i \eta_2 + v_2}{\sqrt{2}}
\end{array}
\right) \;.\label{eq:vevexpansion}
\eeq
The mass matrices are obtained from the terms bilinear in the Higgs
fields in the potential. Due to charge and CP conservation they decompose into $2
\times 2$ matrices ${\cal M}_S$, ${\cal M}_P$ and ${\cal M}_C$ for the
neutral CP-even, neutral CP-odd and charged Higgs sector. They are
diagonalised by the following orthogonal transformations
\beq
\left( \begin{array}{c} \rho_1 \\ \rho_2 \end{array} \right) &=&
R(\alpha) \left( \begin{array}{c} H \\ h \end{array} \right)  \; , \label{eq:diagHh} \\
\left( \begin{array}{c} \eta_1 \\ \eta_2 \end{array} \right) &=&
R(\beta) \left( \begin{array}{c} G^0 \\ A \end{array} \right)  \;
, \label{eq:diagGA} \\
\left( \begin{array}{c} \phi_1^\pm \\ \phi^\pm_2 \end{array} \right) &=&
R(\beta) \left( \begin{array}{c} G^\pm \\ H^\pm \end{array}
\right) \label{eq:diagGHpm}
\;.
\eeq
This leads to the physical Higgs states, a neutral light CP-even, $h$, a neutral heavy
CP-even, $H$, a neutral CP-odd, $A$, and two charged Higgs bosons, 
$H^\pm$. By definition, $m_{h} < m_H$. 
The massless pseudo-Nambu-Goldstone bosons $G^\pm$ and $G^0$
are absorbed by the longitudinal components of the massive gauge
bosons, the charged $W^\pm$ and the $Z$ boson, respectively. The
rotation matrices are given in terms of the mixing angles $\vartheta = \alpha$ and
$\beta$, respectively, and read
\beq
R(\vartheta) = \left( \begin{array}{cc} \cos \vartheta & - \sin
    \vartheta \\ \sin \vartheta & \cos \vartheta \end{array} \right) \;.
\eeq
The mixing angle $\beta$ is related to the two VEVs as
\beq
\tan \beta = \frac{v_2}{v_1} \;, \label{eq:tanbetadef}
\eeq
with $v_1^2 + v_2^2 = v^2 = 1/(\sqrt{2} G_F) \approx (246 \mbox{ GeV})^2$. The mixing angle
$\alpha$ is given by
\beq
\tan 2\alpha = \frac{2 ({\cal M}_S)_{12}}{({\cal M}_S)_{11}-({\cal M}_S)_{22}} \;,
\eeq 
where $({\cal M}_S)_{ij}$ ($i,j=1,2$) denote the matrix elements of the
neutral CP-even scalar mass matrix ${\cal M}_S$. Introducing 
\beq
M^2 \equiv \frac{m_{12}^2}{s_\beta c_\beta}
\eeq
we obtain \cite{Kanemura:2004mg}
\beq
\tan 2\alpha = \frac{s_{2\beta} (M^2- \lambda_{345} v^2)}{c_\beta^2
  (M^2-\lambda_1 v^2) -s_\beta^2 (M^2-\lambda_2 v^2)}
\;,  \label{eq:alphadef}
\eeq
in terms of the abbreviation
\beq
\lambda_{345} \equiv \lambda_3 + \lambda_4 + \lambda_5 
\eeq
and using the short-hand notation $s_x \equiv \sin x$ etc.

In the minimum of the potential, the following conditions have
to be fulfilled,
\beq
\left\langle\frac{\partial V}{\partial \Phi_1}\right\rangle = 
\left\langle\frac{\partial V}{\partial \Phi_2}\right\rangle = 0 \;, \label{eq:tadcond}
\eeq
where the brackets denote the vacuum expectation values. This results in the two equations
\beq
m_{11}^2 &=&  m_{12}^2 \frac{v_2}{v_1} - \frac{\lambda_1 v_1^2}{2}
- \frac{\lambda_{345} v_2^2}{2} \;,\label{eq:tad1} \\
m_{22}^2 &=&  m_{12}^2 \frac{v_1}{v_2} - \frac{\lambda_2 v_2^2}{2}
- \frac{\lambda_{345} v_1^2}{2} \;. \label{eq:tad2}
\eeq
Exploiting the minimum conditions of the potential, we use the
following set of independent input parameters of the model, 
\beq
m_h,\; m_H,\; m_A,\; m_{H^\pm},\; m_{12}^2,\; \alpha,\;
\tan\beta, \; v \;.
\eeq
In this work we choose a
benchmark point of the 2HDM type I, in which the couplings of the two
Higgs doublets to the up- and down-type fermions are equal.
The benchmark point of the 2HDM type I that we use in our numerical
analysis is given by the following set of input parameters
\begin{eqnarray}
\begin{array}{lcllcl}
m_h &=& 125.09 \mbox{ GeV}, \qquad & m_H &=& 134.817 \mbox{ GeV}, \\
m_A &=& 134.711 \mbox{ GeV},  \qquad & m_{H^\pm} &=& 161.5 \mbox{ GeV}, \\
m_{12}^2 &=& 4305 \mbox{ GeV}^2,\qquad & \alpha &=& -0.102, \\ 
\tan\beta &=& 3.759, \qquad & v&=& 246.22 \mbox{ GeV}\;.           
\end{array}
\label{eq:benchmarkpoint}
\end{eqnarray} 
It fulfils all relevant theoretical and experimental constraints. For
a description of the constraints, see Ref.~\cite{Abouabid:2021yvw}.

\section{Calculation}
\label{sec:calculation}

\subsection{\it Partonic leading order cross section}

\begin{figure*}[hbt]
  \centering
  \includegraphics[scale=0.22]{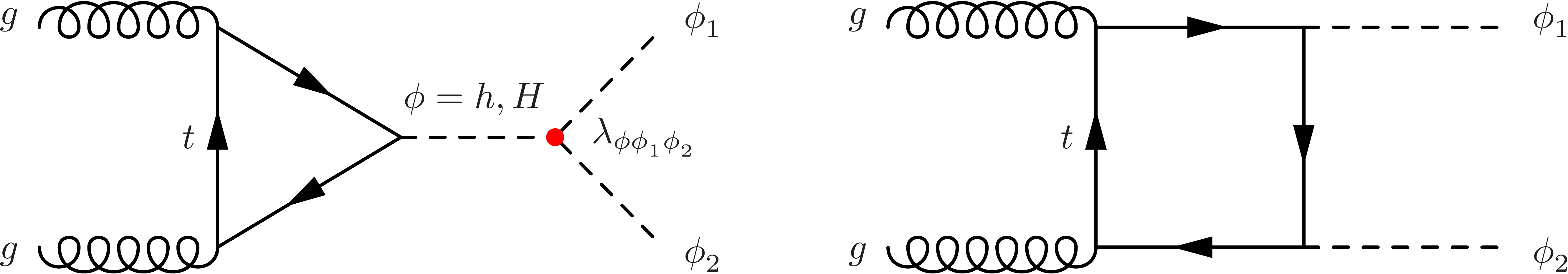}
  \caption[]{\it
    Generic one-loop diagrams for LO Higgs-boson pair
    production via gluon fusion, $gg\to\phi_1^{}\phi_2^{}$, in the 2HDM 
    type I. The contribution from triple Higgs couplings is
    marked in red. Note that $\phi_1^{}\phi_2^{} = hH$ or $AA$.
  }
  \label{fg:lodiags}
\end{figure*}

As we work in the 2HDM type I, we are dominated by the top-quark loop
contributions so that we neglect the bottom-quark loops as well as
light-quark loops. Note that while we work in the 2HDM type I, we could
apply our approximation to the 2HDM (with natural flavour conservation)
of any type as long as we work at low $\tan\beta$ values, as the
top-quark Yukawa coupling is the same in all 2HDM types. In particular
we could apply our approximation to the 2HDM type II and even to the
MSSM as long as the squark contributions can be suppressed, which is the
case for squark mass above 400 GeV~\cite{Dawson:1998py}. This is
typically the case in current MSSM fits to data \cite{GAMBIT:2017snp,
GAMBIT:2017zdo, Bagnaschi:2017tru, Bagnaschi:2018igf, Arbey:2021jdh}.
The leading-order (LO) diagrams for $hH$ and $AA$ production, as
depicted in Fig.~\ref{fg:lodiags} include triangle diagrams, involving a
light and heavy CP-even Higgs $h,H$ propagator coupled to the
final-state Higgs bosons with various triple Higgs couplings, and box
diagrams with two Yukawa couplings. Note, that we focus here on the
production of a mixed CP-even and a pure CP-odd Higgs pair. The analytical 
results and the numerical method for LO and NLO QCD $hh$ and $HH$ 
production can be derived from the SM results
\cite{Glover:1987nx,Plehn:1996wb,Borowka:2016ehy,Borowka:2016ypz,Baglio:2018lrj,Baglio:2020ini}
by simple adjustments of the involved Yukawa and trilinear Higgs
self-couplings as well as the sum over Higgs-boson propagators.
It should be noted that for larger Higgs masses, as e.g. for $HH$ production, 
the top-mass effects and the associated mass and scheme uncertainties 
will be larger than for an SM Higgs mass of 125 GeV.

We follow the conventions of Ref.~\cite{Dawson:1998py} and decompose
the cross section into scalar form factors after the application of
two tensor projectors on the matrix elements. The partonic cross
section $\hat{\sigma}(gg\to \phi_1\phi_2)$, with $\phi_1\phi_2 = hH$
or $AA$, can be written as
\begin{align}
  \hat{\sigma}_{\rm LO} = {\cal S}~
  \frac{G_F^2 \alpha_s^2(\mu_R^2)}{256\,\left(2\pi\right)^3}
     \int_{\hat{t}_-}^{\hat{t}^+} d\hat{t}\, &
  \Big[
  \left|\left(C_\triangle^{h}+C_\triangle^{H}\right) F_\triangle^{} +
  C_\square F_\square^{} \right|^2 \nonumber\\
 & + \left| C_\square G_\square^{} \right|^2\Big],
  \label{eq:xsLO}
\end{align}
where $G_F=1.1663787\cdot 10^{-5}\,\mathrm{GeV^{-2}}$ is the Fermi
constant, $\alpha_s(\mu_R^2)$ is the strong coupling constant
evaluated at the renormalisation scale $\mu_R$, and the Mandelstam
variables $\hat{s}$ and $\hat{t}$ are given by
\begin{align}
  \hat{s}
  & = Q^2 = m_{\phi_1\phi_2}^2,\nonumber\\
  \hat{t}
  & = -\frac12 \left[ Q^2 - m_{\phi_1}^2 - m_{\phi_2}^2 -
    \sqrt{\lambda\left(Q^2_{},m_{\phi_1}^2, m_{\phi_2}^2\right)}\, \cos\theta\right],
    \label{eq:mandelstam}
\end{align}
with the scattering angle $\theta$ in the partonic c.m. system and
where $m_{\phi_1}^{}$ and $m_{\phi_2}^{}$ are the Higgs boson masses,
i.e.~either $m_{\phi_1}^{}=m_h^{}$ and $m_{\phi_2}^{}=m_H^{}$ or
$m_{\phi_1}^{}=m_{\phi_2}^{} = m_A^{}$. The variable
$m_{\phi_1\phi_2}$ denotes the invariant Higgs-pair mass. The factor ${\cal S}$ is a
symmetry factor, ${\cal S}=1/2$ for $AA$ production and ${\cal S}=1$ for $hH$
production. The K\"{a}llen function $\lambda$ is given by
\begin{align}
  \lambda(x,y,z) = (x-y-z)^2_{} - 4 y z.
\end{align}
The integrations limits read as
\begin{align}
  \hat{t}_\pm = -\frac12 \left[ Q^2 - m_{\phi_1}^2 -
  m_{\phi_2}^2\mp\sqrt{\lambda\left(Q^2, m_{\phi_1}^2,
  m_{\phi_2}^2\right)}\, \right].
  \label{eq:mandelstam2}
\end{align}
The coefficients $C_\triangle^{h/H}$ contain the triple Higgs
couplings\linebreak $\lambda_{\phi_1\phi_2 h/H}$ and the
reduced Yukawa couplings $g_{h/H}^t$, which are
  given by the 2HDM Yukawa coupling modification
w.r.t. to the SM top-Yukawa coupling, as well as the CP-even Higgs
boson propagators\footnote{We neglect the total Higgs widths
  $\Gamma_h$ and $\Gamma_H$ in this work which are
    both of ${\cal O}$(MeV) for the chosen benchmark point.},
\begin{align}
  C_\triangle^{h/H} = \lambda_{\phi_1\phi_2 h/H}\,\, g_{h/H}^t\,
  \frac{v}{Q^2-m_{h/H}^2} \;.
  \label{eq:ctriangles}
\end{align}
The coefficient $C_\square$ contains only
reduced Yukawa couplings to the final-state Higgs bosons,
\begin{align}
  C_\square = g_{\phi_1}^t g_{\phi_2}^t.
  \label{eq:cbox}
\end{align}
For the various $\phi_{1,2}$ they are given by
\begin{eqnarray}
g_{h}^t = \cos\alpha/\sin\beta \,, \quad g_{H}^t =
  \sin\alpha/\sin\beta \;, \quad g_A^t = \cot \beta \;.
\end{eqnarray}
In the heavy top-limit (HTL) approximation, the form factors
reduce to
\begin{align}
  F_\triangle^{} = \frac{2}{3}\,a, \,\,\, F_\square^{} = \frac{2}{3},\,\,\, G_\square^{} = 0,
  \label{eq:hHhtlapprox}
\end{align}
with $a=-1$ for $hH$ production and $a=1$ for $AA$ production. The
full $m_t$-dependence at LO can be found in
Refs.~\cite{Glover:1987nx,Plehn:1996wb}.

\subsection{\it Hadronic cross section}

The structure of the NLO QCD corrections is very similar to the SM
case presented in Refs.~\cite{Baglio:2018lrj,Baglio:2020ini}. They
include two-loop virtual corrections to the triangle and box diagrams,
one-particle-reducible diagrams involving two triangle diagrams
connected by a virtual gluon exchange, and one-loop real corrections
involving an extra parton in the final state. The partonic
contributions are then convolved with the parton distributions
functions (PDFs) $f_{i}$ evaluated at the factorisation scale $\mu_F$
in order to obtain the hadronic cross section. The
  parton luminosities $d\mathcal{L}^{ij}/d\tau$ can be defined as
\begin{align}
  \frac{d\mathcal{L}^{i j}}{d\tau} = \int_\tau^1 \frac{dx}{x}
  f_i\left(x,\mu_F\right) f_j\left(\frac{\tau}{x},\mu_F\right),
  \label{eq:pdflumi}
\end{align}
with $\tau=Q^2/s$, $s$ being the hadronic c.m. energy, so that the
NLO hadronic differential cross section with respect to $Q^2$ can be
written as
\begin{align}
  \frac{d\sigma_{\rm NLO}}{dQ^2} = \frac{d\sigma_{\rm LO}}{dQ^2} +
  \frac{d\Delta\sigma_{\rm   virt}}{dQ^2} + \frac{d\Delta
  \sigma_{gg}}{dQ^2} + \frac{d\Delta \sigma_{qg}}{dQ^2} +
  \frac{d\Delta \sigma_{q\bar{q}}}{dQ^2},
  \label{eq:hadronicxs}
\end{align}
with the LO and the virtual and real correction contributions
\begin{align}
  \frac{d\sigma_{\rm LO}}{dQ^2}
  & = \left. \frac{d\mathcal{L}^{gg}}{d\tau}
    \frac{\hat{\sigma}_{\rm LO}\left(Q^2\right)}{s} \right|_{\tau =
\frac{Q^2}{s}},\nonumber\\
  \frac{d\Delta\sigma_{\rm virt}}{dQ^2}
  & = \left. \frac{\alpha_s\left(\mu_R^2\right)}{\pi}
    \frac{d\mathcal{L}^{gg}}{d\tau} \frac{\hat{\sigma}_{\rm LO}\left(Q^2
    \right)}{s}\, C_{\rm virt}\left(Q^2\right) \right|_{\tau = \frac{Q^2}{s}},
    \nonumber\\
  \frac{d\Delta\sigma_{i j}}{dQ^2}
  & = \left. \frac{\alpha_s\left(\mu_R^2\right)}{\pi}\int_{\frac{Q^2}{s}}^1
    \frac{dz}{z^2} \frac{d\mathcal{L}^{i j}}{d\tau}
    \, \frac{\hat{\sigma}_{\rm LO}\left(Q^2\right)}{s}
    C_{i j}(z) \right|_{\tau = \frac{Q^2}{zs}},
    \label{eq:nlodiff}
\end{align}
for $ij = gg$, $\displaystyle\sum_{q,\bar{q}} qg$, and
$\displaystyle\sum_q q\bar{q}$, $z=Q^2/\tau s$, and the variable $\tau$
is restricted to $\tau > \tau_0 =
\left(m_{\phi_1} + m_{\phi_2}\right)_{}^2/s$. We include five external
massless quark flavours. The coefficients $C_{virt}$ of the virtual
and $C_{ij}$ of the real corrections in the HTL have been obtained in
Ref.~\cite{Dawson:1998py} and are given by
\begin{align}
 C_{virt} & = \frac{11}{2} + \pi^2 + C^{\infty,\,
            \phi_1 \phi_2}_{\triangle\triangle} +
\frac{33-2N_F}{6} \log\frac{\mu_R^2}{Q^2}, \nonumber \\
 C_{\triangle\triangle} & = \nonumber\\
& \hspace*{-6mm}\Re e~\frac{\int_{\hat t_-}^{\hat t_+} d\hat t \left\{ \left[ c_1 
C_\square (C_\triangle F_\triangle + F_\Box) + c_2 \frac{p_T^2}{\hat t}
C_\square^2 G_\Box \right] + (\hat t \leftrightarrow \hat u) \right\}}
{\int_{\hat t_-}^{\hat t_+} d\hat t \left\{ |C_\triangle F_\triangle +
C_\square F_\Box |^2 + |C_\square G_\Box|^2 \right\}}, \nonumber \\
C^{\infty,\, hH}_{\triangle\triangle} & = \left. C_{\triangle\triangle}
\right|_{c_1 = c_2 = 2/9}, \nonumber \\
C^{\infty,\, AA}_{\triangle\triangle} & = \left. C_{\triangle\triangle}
\right|_{c_1 = - c_2 = -1/2}, \nonumber \\
C_{gg} & = -z P_{gg}(z) \log\frac{\mu_F^2}{\tau s} - \frac{11}{2}
(1-z)^3 \nonumber \\
& \qquad + 6[1+z^4+(1-z)^4] \left(\frac{\log(1-z)}{1-z}\right)_+, \nonumber \\
C_{gq} & = -\frac{z}{2} P_{gq}(z) \log\frac{\mu_F^2}{\tau s (1-z)^2} +
\frac{2}{3} z^2 - (1-z)^2, \nonumber \\
C_{q\bar q} & = \frac{32}{27} (1-z)^3,
    \label{eq:coeffvirt}
\end{align}
where $C^{\infty,\, hH/AA}_{\triangle\triangle}$ denotes the
contribution of the one-particle reducible diagrams in the HTL
with the transverse momentum $p_T^2 = (\hat t\hat u - m_{\phi_1}^2
m_{\phi_2}^2)/Q^2$ involving $\hat u = m_{\phi_1}^2 + m_{\phi_2}^2 -
Q^2 - \hat t$. The functions $P_{gg}(z)$ and $P_{gq}(z)$ are the related
Altarelli-Parisi splitting kernels \cite{Altarelli:1977zs}, given by
\begin{align}
P_{gg}(z) &= 6\left\{ \left(\frac{1}{1-z}\right)_+
+\frac{1}{z}-2+z(1-z) \right\}
\nonumber \\
& \qquad\qquad\qquad\qquad\qquad\qquad + \frac{33-2N_F}{6}\delta(1-z), 
\nonumber \\
P_{gq}(z) &= \frac{4}{3} \frac{1+(1-z)^2}{z},
\end{align}
with $N_F=5$ in our calculation. The cross section $\hat{\sigma}_{\rm
  LO}(Q^2)$ is calculated in the full theory, i.e. taking into account
the finite top-quark mass at the integrand-level. The total cross
section can be obtained after a final integration over $Q^2$ between
the threshold $\left(m_{\phi_1}+m_{\phi_2}\right)_{}^2$ and the
hadronic c.m. energy $s$.

\subsection{\it Virtual corrections}

\begin{figure*}[hbt]
  \centering
\vspace*{-5cm}

\hspace*{-1.5cm}  \includegraphics[scale=1.0]{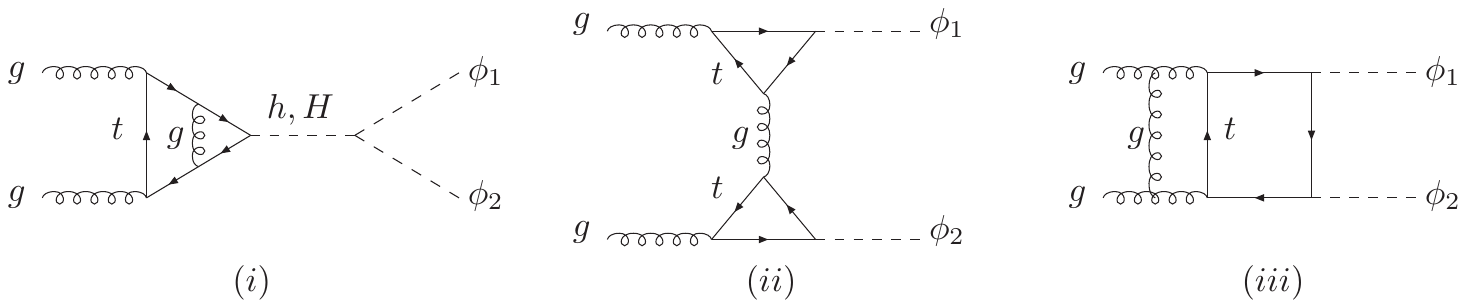}
\vspace*{-21cm}

  \caption[]{\it
    Generic two-loop diagrams for Higgs pair
    production via gluon fusion, $gg\to\phi_1^{}\phi_2^{}~(\phi_1 \phi_2
= hH, AA)$: {\it (i)}
two-loop triangle diagrams, {\it (ii)} one-particle reducible diagrams,
{\it (iii)} two-loop box diagrams.
  }
  \label{fg:nlodiags}
\end{figure*}
Three generic types of diagrams contribute to the virtual
corrections {\it cf.}~Fig.~\ref{fg:nlodiags}:
(i) two-loop triangle diagrams involving the light and heavy scalar
Higgs bosons in the s-channel propagators, (ii) one-particle reducible
diagrams emerging from two triangular top loops coupling to a single
external Higgs boson that are connected by t-channel gluon exchange and
(iii) two-loop box diagrams. The diagrams of class (i) consist of
off-shell single scalar Higgs production dressed with the trilinear
Higgs vertex. The relative QCD corrections coincide with the NLO QCD
corrections to scalar Higgs boson production with mass $Q$ and can thus
be adopted from the single-Higgs calculation \cite{Graudenz:1992pv,
Spira:1995rr, Harlander:2005rq, Anastasiou:2009kn, Aglietti:2006tp}. The
diagrams of class (ii) define the coefficients $c_1, c_2$ in
Eq.~(\ref{eq:coeffvirt}). The analytical expressions of the coefficients $c_1, c_2$ of the
one-particle reducible contributions can be obtained from the
corresponding Higgs decay widths of $\phi \to Z\gamma~(\phi = h,H,A)$
\cite{Cahn:1978nz, Bergstrom:1985hp, Gamberini:1987sv} with the
corresponding adjustments of the involved couplings.
The full top-mass dependence of $c_1, c_2$ is given by\footnote{In the case of different pseudoscalar
Higgs bosons as in more extended Higgs sectors, the coefficient reads
$c_1=-c_2 = -2 I_2(\tau_{A_1},\lambda_{\hat t})
I_2(\tau_{A_2},\lambda_{\hat t})$ where $\tau_{A_k} = 4
m_t^2/m_{A_k}^2~(k=1,2)$ for the two pseudoscalars $A_{1,2}$.}
\begin{align}
c_1 & = c_2 = 2
\left[ I_1(\tau_h,\lambda_{\hat t}) - I_2(\tau_h,\lambda_{\hat t})
\right] \nonumber \\
& \qquad\;\;\;\; \times
\left[ I_1(\tau_H,\lambda_{\hat t}) - I_2(\tau_H,\lambda_{\hat t})
\right]
\quad
\mbox{for $\phi_1\phi_2 = hH$} \nonumber \\
c_1 & = -c_2 = -2
\left[ I_2(\tau_A,\lambda_{\hat t}) \right]^2
\qquad\qquad \mbox{for $\phi_1\phi_2 = AA$} \nonumber \\
I_1(\tau,\lambda) & = \frac{\tau\lambda}{2(\tau-\lambda)} +
\frac{\tau^2\lambda^2}{2(\tau-\lambda)^2} \left[ f(\tau) - f(\lambda)
\right] \nonumber \\
& \qquad + \frac{\tau^2\lambda}{(\tau-\lambda)^2} \left[ g(\tau)
- g(\lambda) \right], \nonumber \\
I_2(\tau,\lambda) & = - \frac{\tau\lambda}{2(\tau-\lambda)}\left[
f(\tau) - f(\lambda) \right]
\end{align}
with $\tau_\phi = 4m_t^2/m_\phi^2~(\phi=h,H,A)$ and $\lambda_{\hat t} =
4m_t^2/\hat t$. The generic loop functions are given by
\begin{align}
f(\tau) & = \left\{ \begin{array}{ll} \displaystyle \arcsin^2
\frac{1}{\sqrt{\tau}} & \tau \ge 1 \\ \displaystyle - \frac{1}{4} \left[
\log \frac{1+\sqrt{1-\tau}} {1-\sqrt{1-\tau}} - i\pi \right]^2 & \tau <
1 \end{array} \right. \nonumber \\
g(\tau) & = \left\{ \begin{array}{ll} \displaystyle \sqrt{\tau-1}~
\arcsin \frac{1}{\sqrt{\tau}} & \tau \ge 1 \\ \displaystyle
\frac{\sqrt{1-\tau}}{2} \left[ \log \frac{1+\sqrt{1-\tau}}
{1-\sqrt{1-\tau}} - i\pi \right] & \tau < 1 \end{array} \right.
\end{align}
These expressions approach the HTL values given in
Eq.~(\ref{eq:coeffvirt}).

The involved part of our calculation is the two-loop box diagrams of
type (iii). We have used the same method as in
Refs.~\cite{Baglio:2018lrj, Baglio:2020ini, Baglio:2020wgt},
i.e.~we have performed a Feynman parametrisation, end-point subtractions and
the subtraction of special infrared terms to allow for a clean separation
of the ultraviolet and infrared singularities. For the stabilisation of
the 6-dimensional Feynman integrals we have
    applied integrations by
parts to reduce the powers of the singular denominators and performed the
integrations with a small imaginary part of the virtual top mass. In
order to arrive at the narrow-width approximation for the virtual top
mass, we have used Richardson extrapolations \cite{richardson} along
the lines of our SM calculation of Refs.~\cite{Baglio:2018lrj,
Baglio:2020ini}. However, here we needed to extend
the calculation for scalar Higgs-boson pairs to the case of {\it
different} final-state Higgs masses. For the calculation of pseudoscalar
Higgs-boson pairs, we have used a naive anti-commuting $\gamma_5$ matrix
at the pseudoscalar vertices, since only even numbers of $\gamma_5$
contribute to the (${\cal CP}$-even) virtual corrections diagram by
diagram. For this case, we have used the
same projectors as in the 
double-scalar case, since the contributing tensor structures are the
same. Since each individual two-loop box diagram is singular for the
$\hat t$ integration, we have applied a technical cut at the integration
boundaries and included a suitable substitution to stabilise this
integration for each diagram. We have checked explicitly that our
results do not depend on this technical cut.

The top mass has been renormalised in both the on-shell \linebreak
scheme and in the $\overline{\rm MS}$ scheme. The on-shell scheme
predictions are our default central predictions while the
$\overline{\rm MS}$ scheme predictions are used to calculate the
top-quark scale and scheme uncertainties, see below. The strong
coupling constant is renormalised in the $\overline{\rm MS}$ scheme
with 5 active flavours. We have obtained finite results for
the virtual corrections by subtracting the HTL results as in the SM
case so that we end
up effectively calculating the NLO mass effects only. To obtain the
final hadronic differential cross section, we have
added back the HTL results 
calculated with {\tt HPAIR}\footnote{The program can be downloaded at
\href{http://tiger.web.psi.ch/hpair/}{http://tiger.web.psi.ch/hpair/}.}.
The calculation of each two-loop box diagram has been performed
independently at least twice with different Feynman parametrisations
and we have obtained full agreement within the numerical precision.

\subsection{\it Real corrections}

The calculation of the finite mass effects in the real corrections,
$\Delta \sigma_{ij}^{\rm mass} = \Delta \sigma_{ij} - \Delta
\sigma_{ij}^{\rm HTL}$, follows closely the method described in
Refs.~\cite{Baglio:2018lrj,Baglio:2020ini} for the SM case. The HTL
contributions are calculated again with the program {\tt HPAIR} while
the partonic mass effects are obtained as
\begin{align}
 d\Delta\hat{\sigma}_{ij}^{\rm mass} = d\Delta\hat{\sigma}_{ij} -
  d\hat{\sigma}_{\rm LO}(\tilde p_i) \frac{d\Delta\hat{\sigma}_{ij}^{\rm
  HTL}(p_i)}{d\hat{\sigma}_{\rm LO}^{\rm HTL}(\tilde p_i)},
\end{align}
where the exact four-momenta $p_i$ are mapped onto LO sub-space
four-momenta $\tilde p_i$ following Ref.~\cite{Catani:1996vz}.

The HTL matrix elements have been calculated analytically, while the full
one-loop matrix elements have been obtained by two different
methods. They have been generated using {\tt
  FeynArts}~\cite{Hahn:2000kx} and {\tt FormCalc}~\cite{Hahn:1998yk}
on the one hand, and obtained analytically using {\tt
  FeynCalc}~\cite{Shtabovenko:2020gxv} on the other hand. The scalar
one-loop integrals have then been
  calculated numerically using the library
{\tt COLLIER 1.2}~\cite{Denner:2016kdg}. The phase-space has also been parameterised
in two different ways. The two methods agree within
the numerical precision.
 
\section{Numerical results}
\label{sec:numresults}

\begin{figure*}[t!]
  \centering
  \includegraphics[scale=0.63]{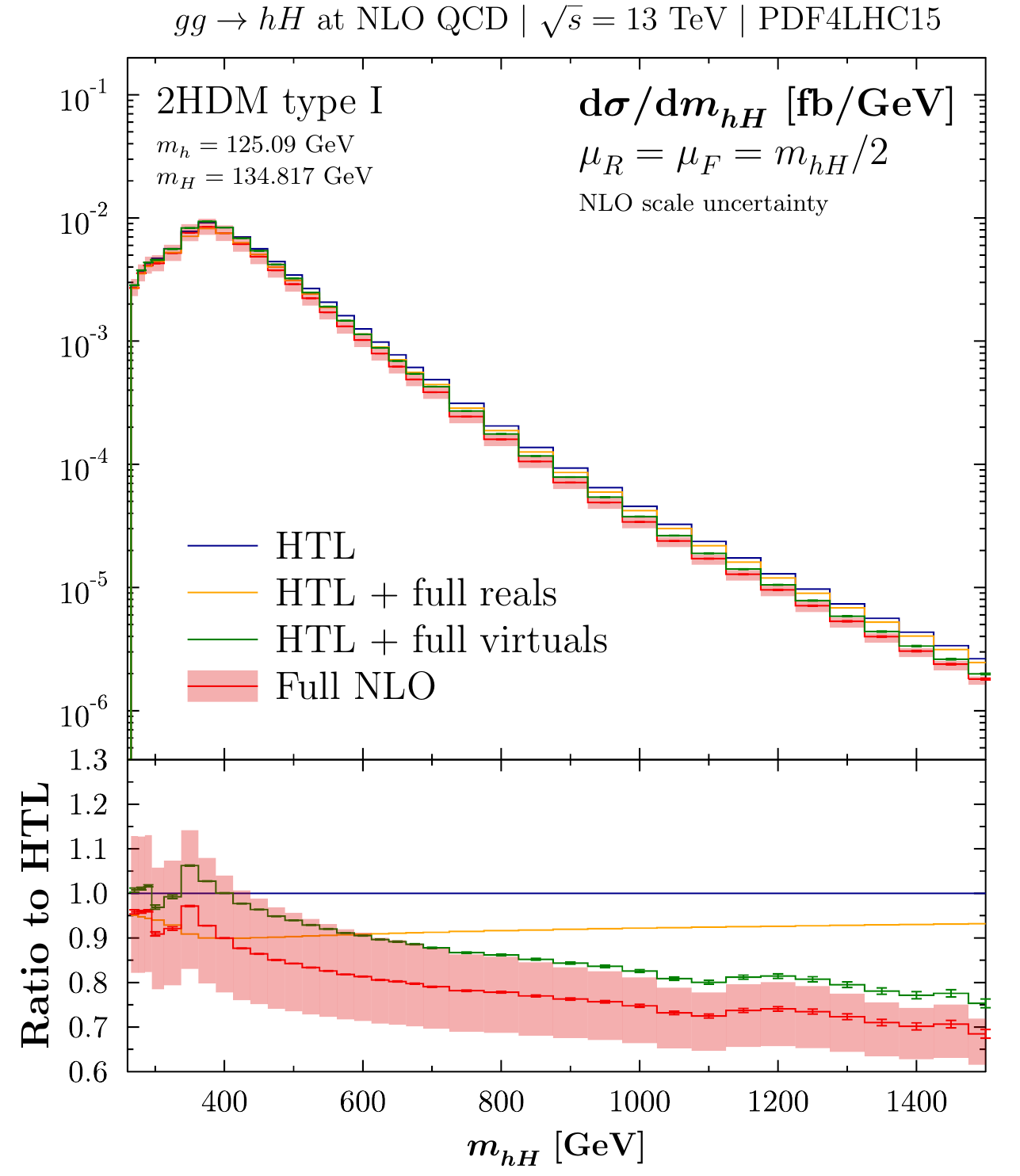}
  \hspace{3mm}
  \includegraphics[scale=0.63]{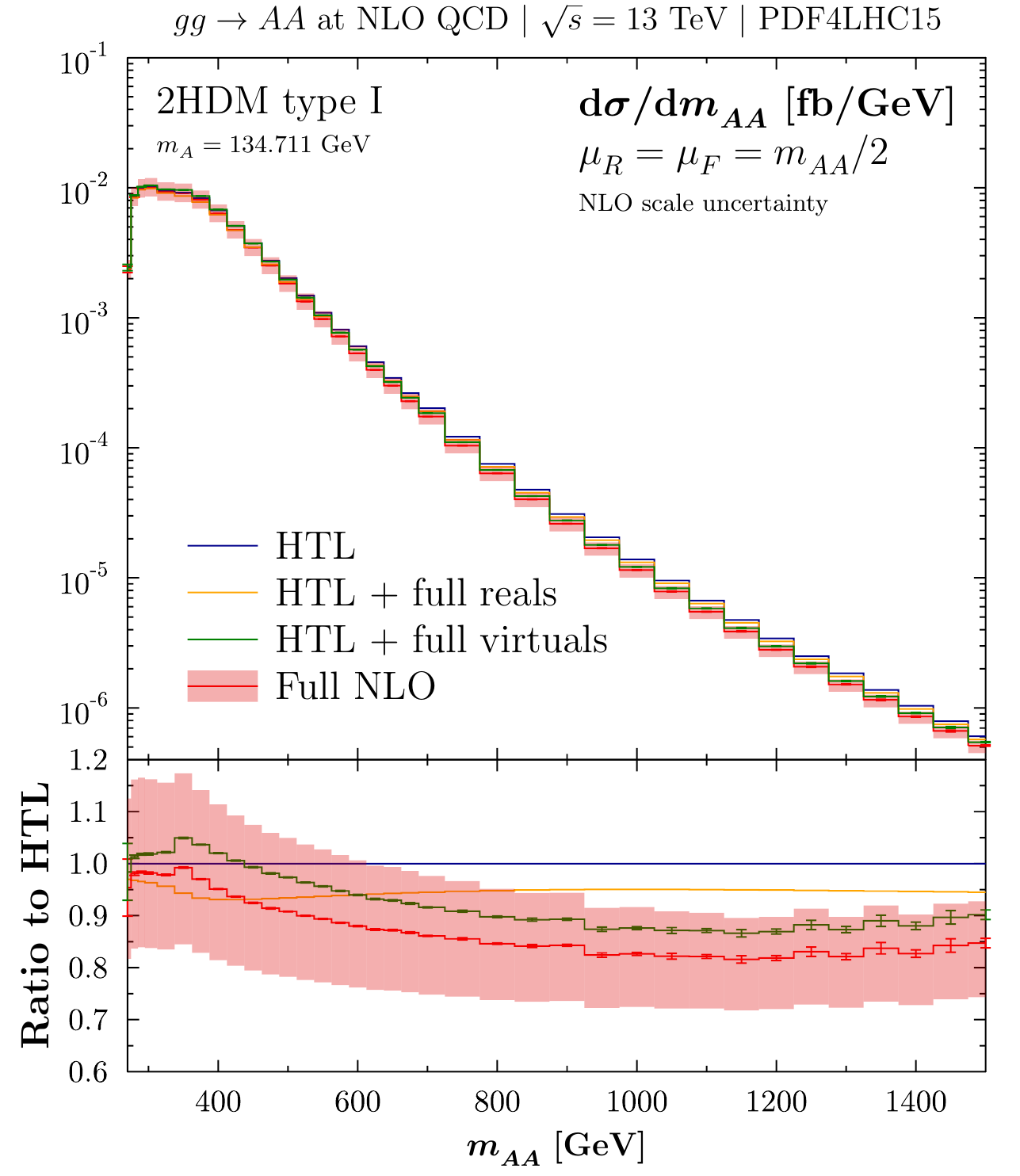}
  \caption[]{Invariant Higgs-pair-mass distributions for Higgs boson
    pair production via gluon fusion at the 13 TeV LHC as a function
    of $Q$ using the {\tt PDF4LHC15} PDF set, in the 2HDM type
    I. Left: CP-even $hH$ production. Right: CP-odd $AA$
    production. In both panels, the Born-improved HTL results (in
    blue), HTL results including the full real corrections (in
    yellow), HTL results including the full virtual corrections (in
    green, including the numerical error), and the full NLO QCD
    results (in red, including the numerical error) are depicted. The
    inserts below display the ratio to the NLO HTL result for the
    different calculations. The red band indicates the renormalisation
    and factorisation scale uncertainties for the results
    including the full NLO QCD corrections.}
  \label{fig:distrib13}
\end{figure*}

We present our numerical results at a hadron $pp$ collider for c.m.~energies of
$\sqrt{s}=13$ and 14~TeV (LHC energies), $\sqrt{s}=27$~TeV
(high-energy variant of the LHC, the HE-LHC), and $\sqrt{s}=100$~TeV
(FCC energy). We use $m_t=172.5$~GeV for the on-shell top-quark
mass. We have performed the calculation using the NLO PDF set {\tt
PDF4LHC15}~\cite{Butterworth:2015oua} as implemented in the {\tt
LHAPDF-6} library~\cite{Buckley:2014ana}. Our central scale choice is
$\mu_R=\mu_F=\mu_0 = Q/2$, and $\alpha_s(M_Z^2)$ is set according
to the chosen PDF set, with an NLO running in the
five-flavour scheme. As done also in the SM
calculation~\cite{Baglio:2018lrj,Baglio:2020ini}, we have used the
narrow-width approximation for the top quark. We use the 2HDM
benchmark scenario given in Eq.~(\ref{eq:benchmarkpoint}).

\begin{figure*}[t!]
  \centering
  \includegraphics[scale=0.63]{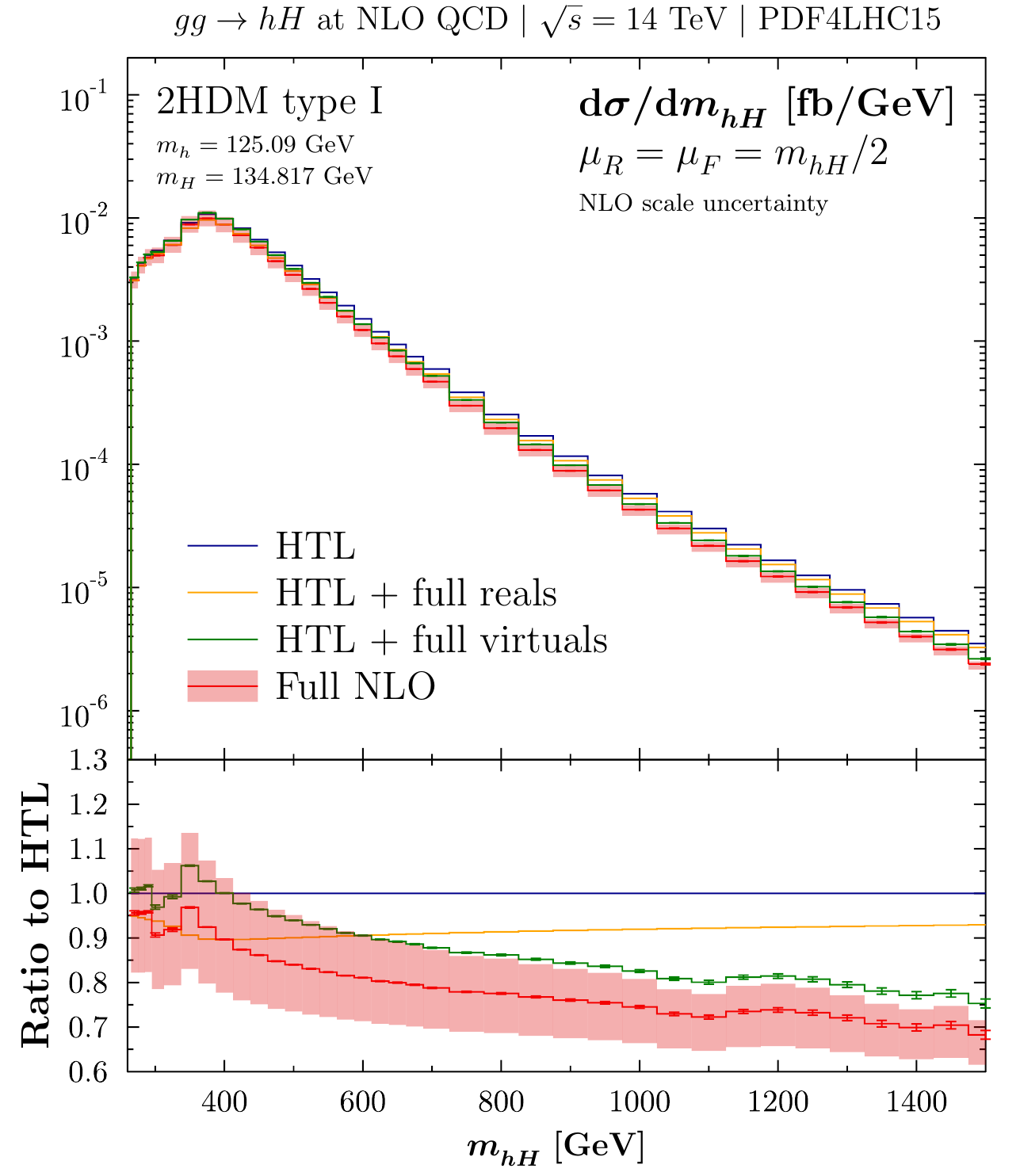}
  \hspace{3mm}
  \includegraphics[scale=0.63]{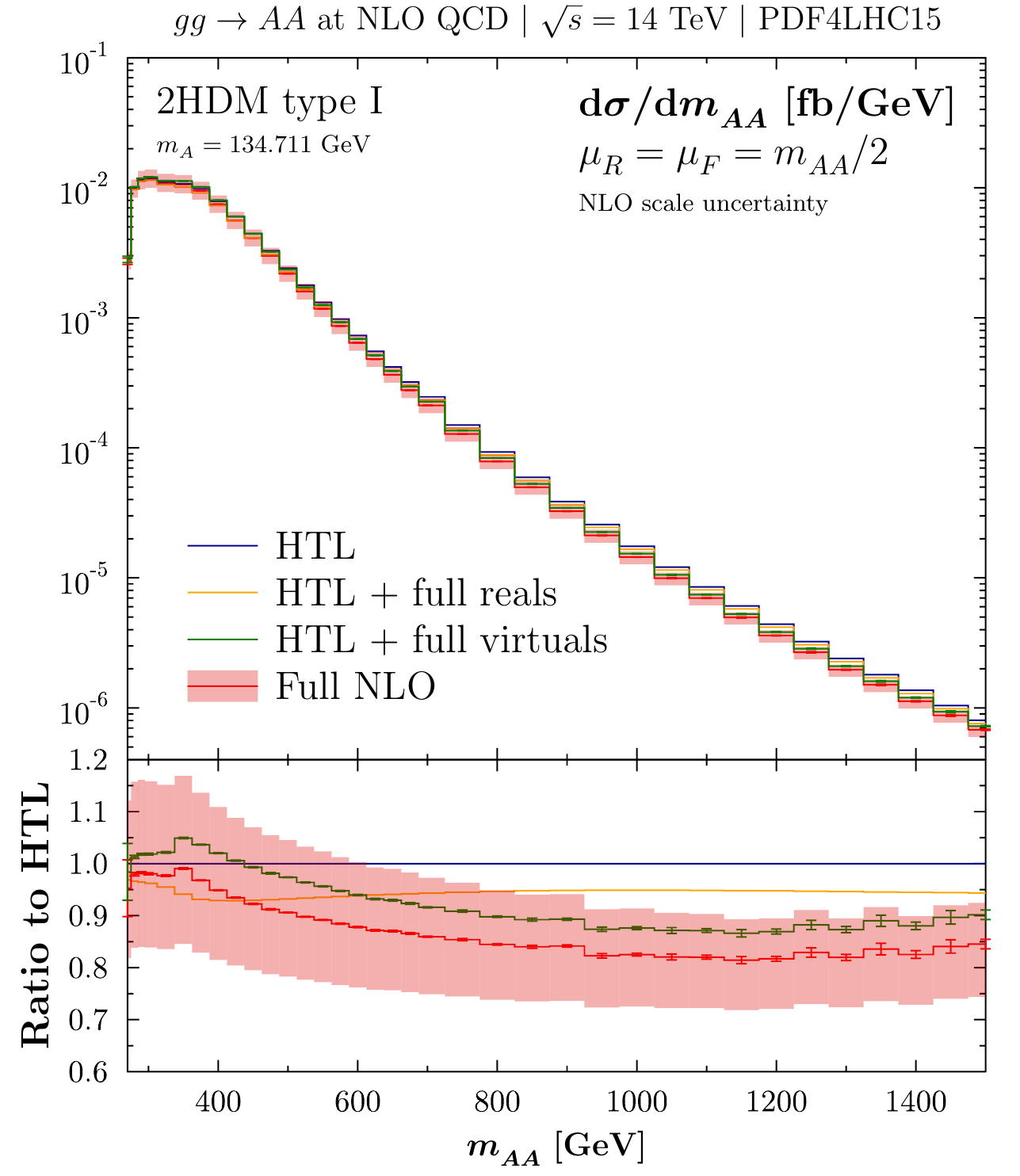}
  \caption[]{Same as in Fig.~\protect\ref{fig:distrib13} but for
    $\sqrt{s}=14~\mathrm{TeV}$.}
  \label{fig:distrib14}
\end{figure*}

\begin{figure*}[t!]
  \centering
  \includegraphics[scale=0.63]{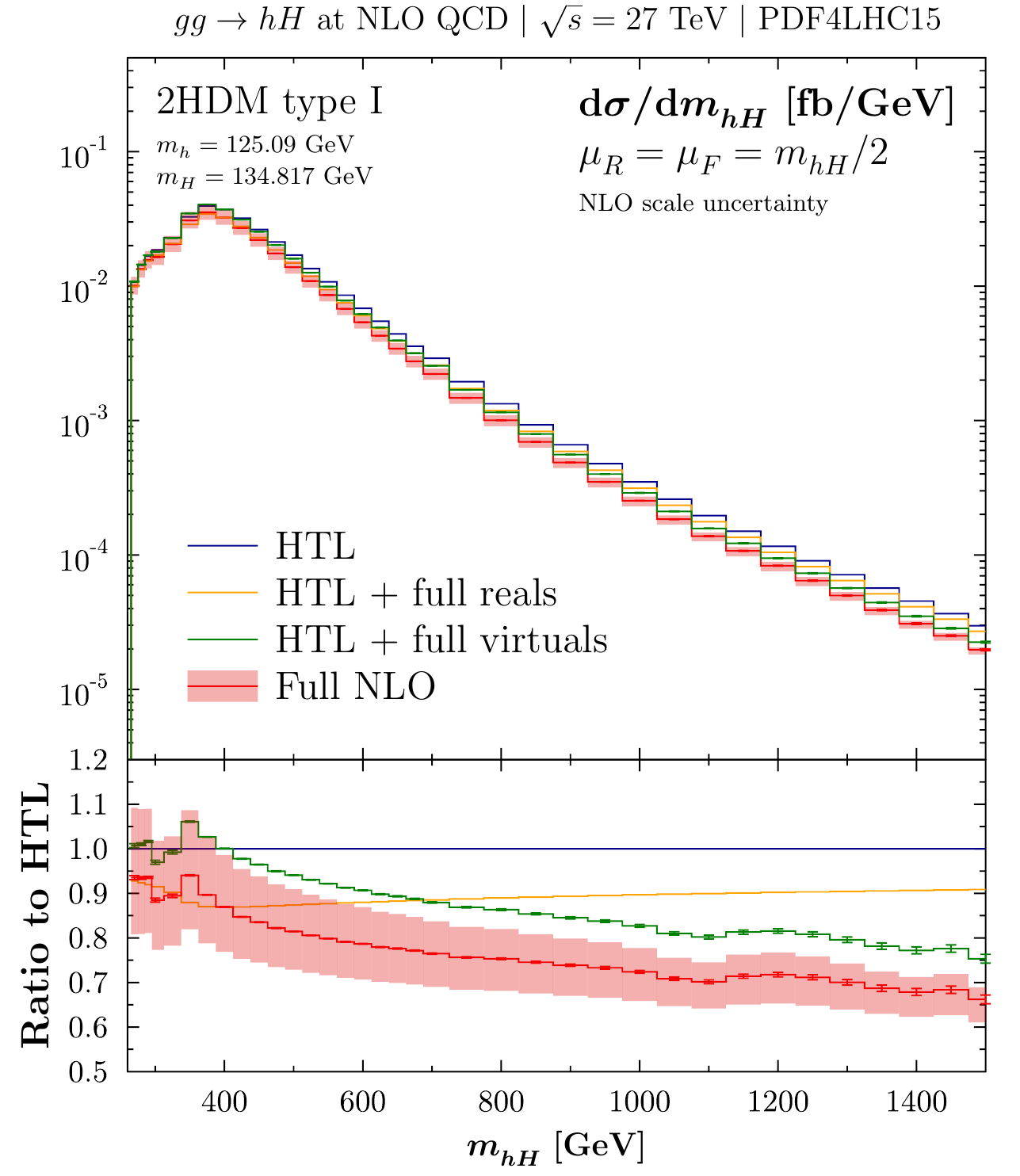}
  \hspace{3mm}
  \includegraphics[scale=0.63]{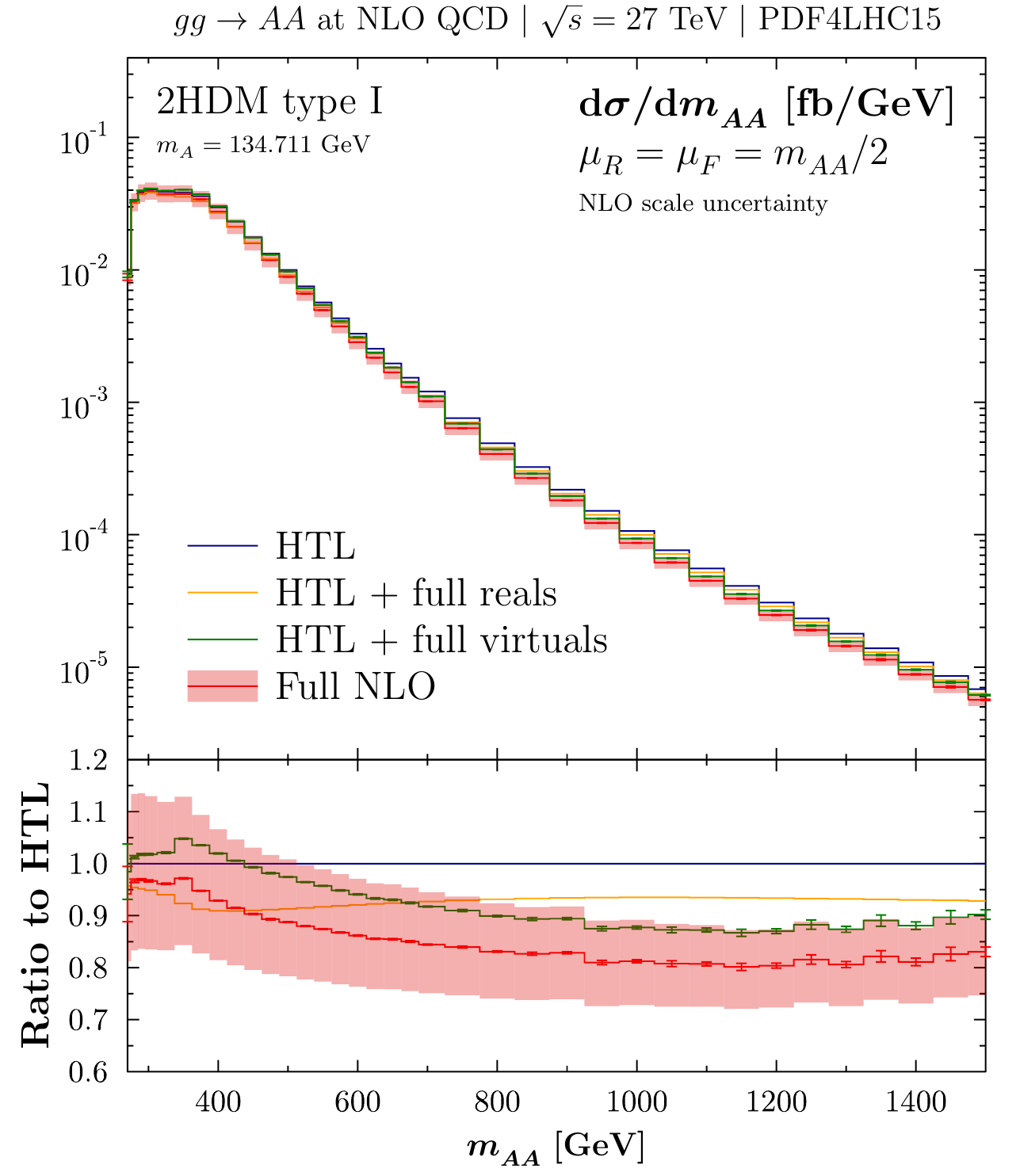}
  \caption[]{Same as in Fig.~\protect\ref{fig:distrib13} but for $\sqrt{s}=27~\mathrm{TeV}$.}
  \label{fig:distrib27}
\end{figure*}

\begin{figure*}[t!]
  \centering
  \includegraphics[scale=0.63]{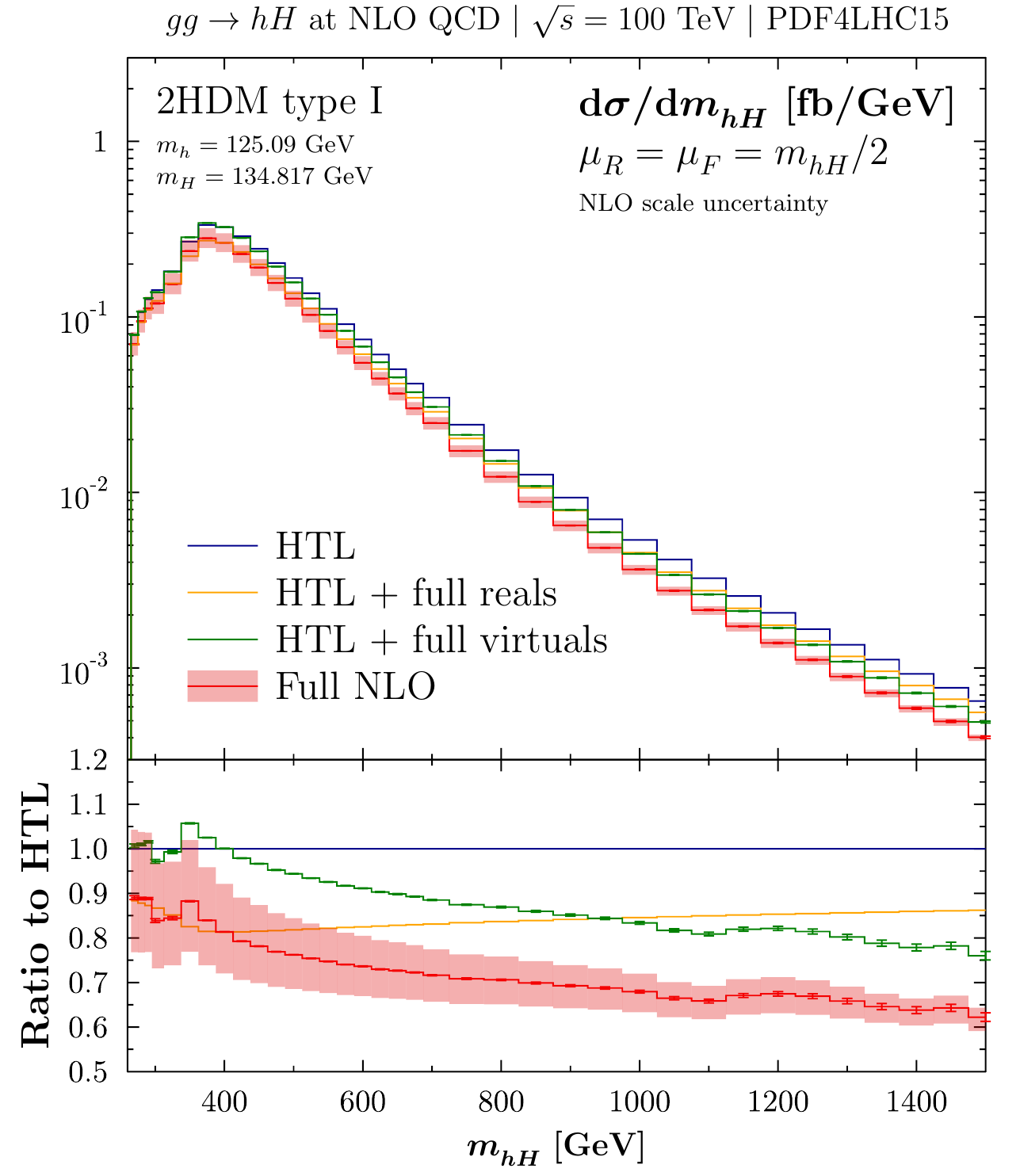}
  \hspace{3mm}
  \includegraphics[scale=0.63]{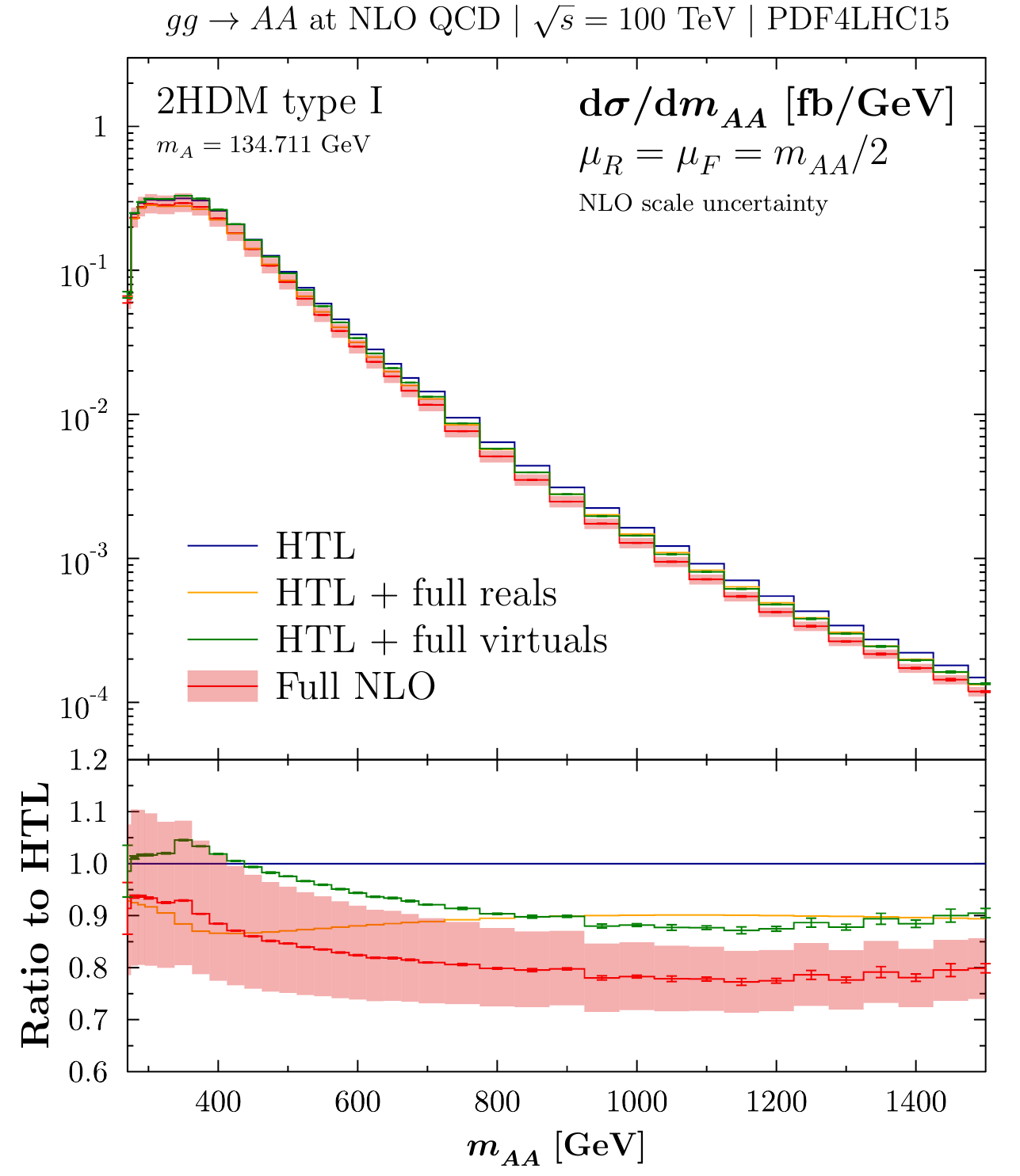}
  \caption[]{Same as in Fig.~\protect\ref{fig:distrib13} but for $\sqrt{s}=100~\mathrm{TeV}$.}
  \label{fig:distrib100}
\end{figure*}

We have calculated a grid of $Q$-values from $Q=259.907$
$(269.422)~\mathrm{GeV}$, for $hH$ production (for $AA$ production), to
$Q=1500~\mathrm{GeV}$, so that we obtain the invariant Higgs-pair-mass
distributions depicted in Fig.~\ref{fig:distrib13} for $hH$ production
(left) and $AA$ production (right), for the LHC at 13 TeV. The results
at 14 TeV are shown in Fig.~\ref{fig:distrib14}, while the results for
the HE-LHC are shown in Fig.~\ref{fig:distrib27} and the results for the
FCC in Fig.~\ref{fig:distrib100}. The full NLO QCD results are displayed
in red, including the numerical errors as well as a band indicating the
renormalisation and factorisation scale uncertainties obtained with a
standard seven-point variation around our central scale choice ({\it
cf.}~Subsec.~\ref{subsec:scale}). The blue line shows the
(Born-improved) HTL prediction, while the yellow line displays the HTL
supplemented by the full mass effects in the real corrections only and
the green line (including numerical errors) the HTL supplemented by the
full mass effects in the virtual corrections only.


The mass effects in the real corrections increase with increasing
c.m. energy both for $hH$ and $AA$ final states. In CP-even $hH$
production, they reach a negative peak at around $Q=400~\mathrm{GeV}$
and are of the order of $-10\%$ at 13 TeV (of the order of $-20\%$ at
100 TeV) before mildly increasing up to around -$6\%$ at
$Q=1500~\mathrm{GeV}$ at 13 TeV ($-14\%$ at 100 TeV). In CP-odd $AA$
production, the behaviour of the mass effects in the real corrections
is slightly different. There is also a negative peak around
$Q=400~\mathrm{GeV}$, of the order of $-8\%$ at 13 TeV ($-14\%$ at 100
TeV), but then it mildly increases before reaching a plateau around
$Q=1000~\mathrm{GeV}$. The mass effects are then practically constant,
about $-6\%$ at 13 TeV ($-11\%$ at 100 TeV). The mass effects in the
virtual corrections are negative at large $Q$ values for both $hH$
and $AA$ final states, as expected by the restoration of partial-wave
unitarity in the high-energy limit. Combined with the mass effects in
the real corrections, the full mass effects reach about $-30\%$ ($-40\%$) at
$Q\simeq 1500~\mathrm{GeV}$ for $hH$ production, at lower
c.m. energies (at 100 TeV),
while the mass effects in the virtual corrections are smaller for $AA$
production, reaching about $-15\%$ ($-20\%$ for $Q \simeq 1500~\mathrm{GeV}$,
at lower c.m. energies (at 100 TeV).
  This is the same behaviour that is
observed in the SM
case~\cite{Borowka:2016ehy,Borowka:2016ypz,Baglio:2018lrj,Baglio:2020ini},
albeit with a smaller correction for $AA$ production. Note
that the mild increase in the mass effects in the virtual corrections
at large $Q$ values for $AA$ production can be attributed to numerical
fluctuations. The most striking difference between CP-even and
CP-odd pair production can be seen around the $t\bar{t}$
threshold and below. There is a distortion of the shape that is distinctly
different from the SM case and also between $hH$ and $AA$
productions, hence discriminating between the two production
channels.

We have also obtained the total cross sections from the differential
distributions, using a numerical integration of $Q$. For $Q$ between
300~GeV and 1500~GeV we have used the trapezoidal method
supplemented by a 
Richardson extrapolation~\cite{richardson} while we use a Simpson's
3/8 rule~\cite{Abramowitz} for $Q$ between \linebreak 270~GeV and 300~GeV and a
simple trapezoid for $Q$ between the threshold and 270~GeV. For the
FCC c.m. energy of 100 TeV we have also included three new $Q$ bins between
1500~GeV and 2500~GeV and add their contribution using a Simpson's
rule. Including the numerical errors on the final decimal number, we
have obtained the following results for the full NLO QCD total cross sections
for $hH$ and $AA$ production in our 2HDM benchmark scenario, using
{\tt PDF4LHC15} PDF sets,
\begin{align}
  13~\mathrm{TeV}:\, \,
  \sigma_{gg\to hH}^{} = 1.592(1)\,\mathrm{fb}, \quad
  \sigma_{gg\to AA}^{} = 1.643(1)\,\mathrm{fb}, \nonumber\\
  14~\mathrm{TeV}:\, \,
  \sigma_{gg\to hH}^{} = 1.876(1)\,\mathrm{fb}, \quad
  \sigma_{gg\to AA}^{} = 1.927(1)\,\mathrm{fb}, \nonumber\\
  27~\mathrm{TeV}:\, \,
  \sigma_{gg\to hH}^{} = 7.036(4)\,\mathrm{fb}, \quad
  \sigma_{gg\to AA}^{} = 7.012(4)\,\mathrm{fb}, \nonumber\\
  100~\mathrm{TeV}:\, \,
  \sigma_{gg\to hH}^{} = 60.49(4)\,\mathrm{fb}, \quad
  \sigma_{gg\to AA}^{} = 58.12(3)\,\mathrm{fb}.
  \label{eq:finalxs}
\end{align}
The corresponding results in the (Born-improved) HTL approximation,
obtained using the same numerical integration of the $Q$ grid, are
\begin{align}
  13~\mathrm{TeV}:\, \,
  \sigma_{gg\to hH}^{\rm HTL} = 1.793\,\mathrm{fb}, \quad
  \sigma_{gg\to AA}^{\rm HTL} = 1.717\,\mathrm{fb},
  \nonumber\\
  14~\mathrm{TeV}:\, \,
  \sigma_{gg\to hH}^{\rm HTL} = 2.120\,\mathrm{fb}, \quad
  \sigma_{gg\to AA}^{\rm HTL} = 2.018\,\mathrm{fb},
  \nonumber\\
  27~\mathrm{TeV}:\, \,
    \sigma_{gg\to hH}^{\rm HTL} = 8.240\,\mathrm{fb}, \quad
  \sigma_{gg\to AA}^{\rm HTL} = 7.504\,\mathrm{fb},
  \nonumber\\
  100~\mathrm{TeV}:\, \,
    \sigma_{gg\to hH}^{\rm HTL} = 76.32\,\mathrm{fb}, \quad
  \sigma_{gg\to AA}^{\rm HTL} = 65.28\,\mathrm{fb}.
  \label{eq:finalxshtl}
\end{align}
The comparison of Eq.~(\ref{eq:finalxs}) with
Eq.~(\ref{eq:finalxshtl}) gives a $\simeq -12\%$ top-mass effect
correction at NLO on the total cross section for $hH$ production at
LHC energies ($\simeq -21\%$ at the 100 TeV FCC), and a $\simeq -5\%$
correction for $AA$ production at LHC energies ($\simeq -11\%$ at the
100 TeV FCC). While the mass effects are of similar size as the SM
Higgs-pair production for CP-even Higgs bosons, they are smaller
for CP-odd Higgs pair production.

\section{Theoretical uncertainties}
\label{sec:errors}

\subsection{\it Factorisation and renormalisation scale uncertainties}
\label{subsec:scale}

We have estimated the factorisation and renormalisation scale
uncertainties using the standard seven-point method. We have varied
both the factorisation scale $\mu_F$ and the renormalisation scale
$\mu_R$ around our central scale choice $\mu_R=\mu_F=Q/2$, by a factor
of two up and down while avoiding the choices leading to the ratio
$\mu_R/\mu_F$ being either greater than two or smaller than
one-half. The maximal and minimal cross sections obtained by this
procedure are then compared to the nominal cross section obtained with
the central scale choice.

We have obtained for the total cross section calculated
with \linebreak {\tt PDF4LHC15} 
parton densities the following scale
uncertainties for CP-even Higgs-pair production $hH$,
\begin{align}
  13~\mathrm{TeV}: \quad
  \sigma_{gg\to hH}^{} = 1.592(1)^{+15.2\%}_{-13.4\%}\,\mathrm{fb}, \nonumber\\
  14~\mathrm{TeV}: \quad
  \sigma_{gg\to hH}^{} = 1.876(1)^{+14.9\%}_{-13.2\%}\,\mathrm{fb}, \nonumber\\
  27~\mathrm{TeV}: \quad
  \sigma_{gg\to hH}^{} = 7.036(4)^{+13.1\%}_{-11.4\%}\,\mathrm{fb}, \nonumber\\
  100~\mathrm{TeV}: \quad
  \sigma_{gg\to hH}^{} = 60.49(4)^{+12.4\%}_{-10.9\%}\,\mathrm{fb},
  \label{eq:hHscaleerrors}
\end{align}
while we have obtained the following results for CP-odd Higgs-pair production
$AA$,
\begin{align}
  13~\mathrm{TeV}: \quad
  \sigma_{gg\to AA}^{} = 1.643(1)^{+17.4\%}_{-14.4\%}\,\mathrm{fb}, \nonumber\\
  14~\mathrm{TeV}: \quad
  \sigma_{gg\to AA}^{} = 1.927(1)^{+17.1\%}_{-14.2\%}\,\mathrm{fb}, \nonumber\\
  27~\mathrm{TeV}: \quad
  \sigma_{gg\to AA}^{} = 7.012(4)^{+15.3\%}_{-12.7\%}\,\mathrm{fb}, \nonumber\\
  100~\mathrm{TeV}: \quad
  \sigma_{gg\to AA}^{} = 58.12(3)^{+15.2\%}_{-12.6\%}\,\mathrm{fb}.
  \label{eq:AAscaleerrors}
\end{align}
The scale uncertainties are similar to what is obtained for SM Higgs
pair
production~\cite{Borowka:2016ehy,Borowka:2016ypz,Baglio:2018lrj,Baglio:2020ini}. They
are slightly larger in $AA$ production than in $hH$ production. We have
also found the following scale dependences for the differential
cross section at 13 TeV for four distinct values of $Q$,
\begin{align}
  \frac{d\sigma(gg\to hH)}{dQ}\Big|_{Q=300~{\rm GeV}} & = 0.004278(2)^{+16.4\%}_{-13.6\%}\,
                                                        \mathrm{fb/GeV},\nonumber\\
  \frac{d\sigma(gg\to hH)}{dQ}\Big|_{Q=400~{\rm GeV}} & = 0.007522(5)^{+15.6\%}_{-13.6\%}\,
                                                        \mathrm{fb/GeV},\nonumber\\
  \frac{d\sigma(gg\to hH)}{dQ}\Big|_{Q=600~{\rm GeV}} & = 0.0010217(9)^{+12.1\%}_{-12.3\%}\,
                                                        \mathrm{fb/GeV},\nonumber\\
  \frac{d\sigma(gg\to hH)}{dQ}\Big|_{Q=1200~{\rm GeV}} & = 0.00000956(6)^{+8.1\%}_{-11.3\%}\,
                                                         \mathrm{fb/GeV},
\end{align}
and
\begin{align}
  \frac{d\sigma(gg\to AA)}{dQ}\Big|_{Q=300~{\rm GeV}} & = 0.01005(2)^{+18.3\%}_{-14.7\%}\,
                                                        \mathrm{fb/GeV},\nonumber\\
  \frac{d\sigma(gg\to AA)}{dQ}\Big|_{Q=400~{\rm GeV}} & = 0.006346(6)^{+17.1\%}_{-14.4\%}\,
                                                        \mathrm{fb/GeV},\nonumber\\
  \frac{d\sigma(gg\to AA)}{dQ}\Big|_{Q=600~{\rm GeV}} & = 0.0005328(7)^{+14.4\%}_{-13.4\%}\,
                                                        \mathrm{fb/GeV},\nonumber\\
  \frac{d\sigma(gg\to AA)}{dQ}\Big|_{Q=1200~{\rm GeV}} & = 0.00000280(2)^{+9.7\%}_{-12.0\%}\,
                                                         \mathrm{fb/GeV}.
\end{align}

\subsection{\it Top-quark scale and scheme uncertainties}
\label{subsec:topquark}

\begin{figure*}[t!]
  \centering
  \includegraphics[scale=0.63]{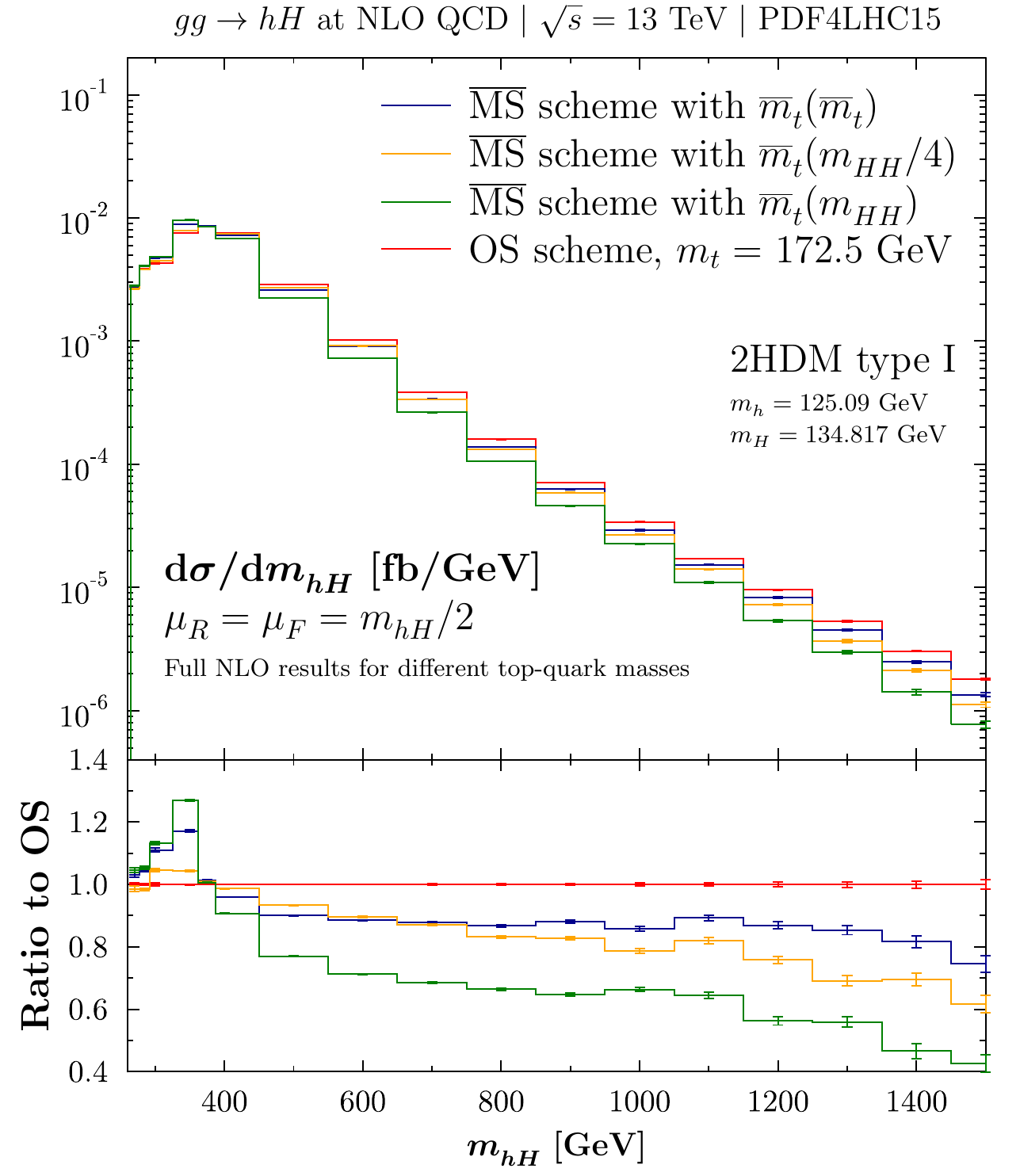}
  \hspace{3mm}
  \includegraphics[scale=0.63]{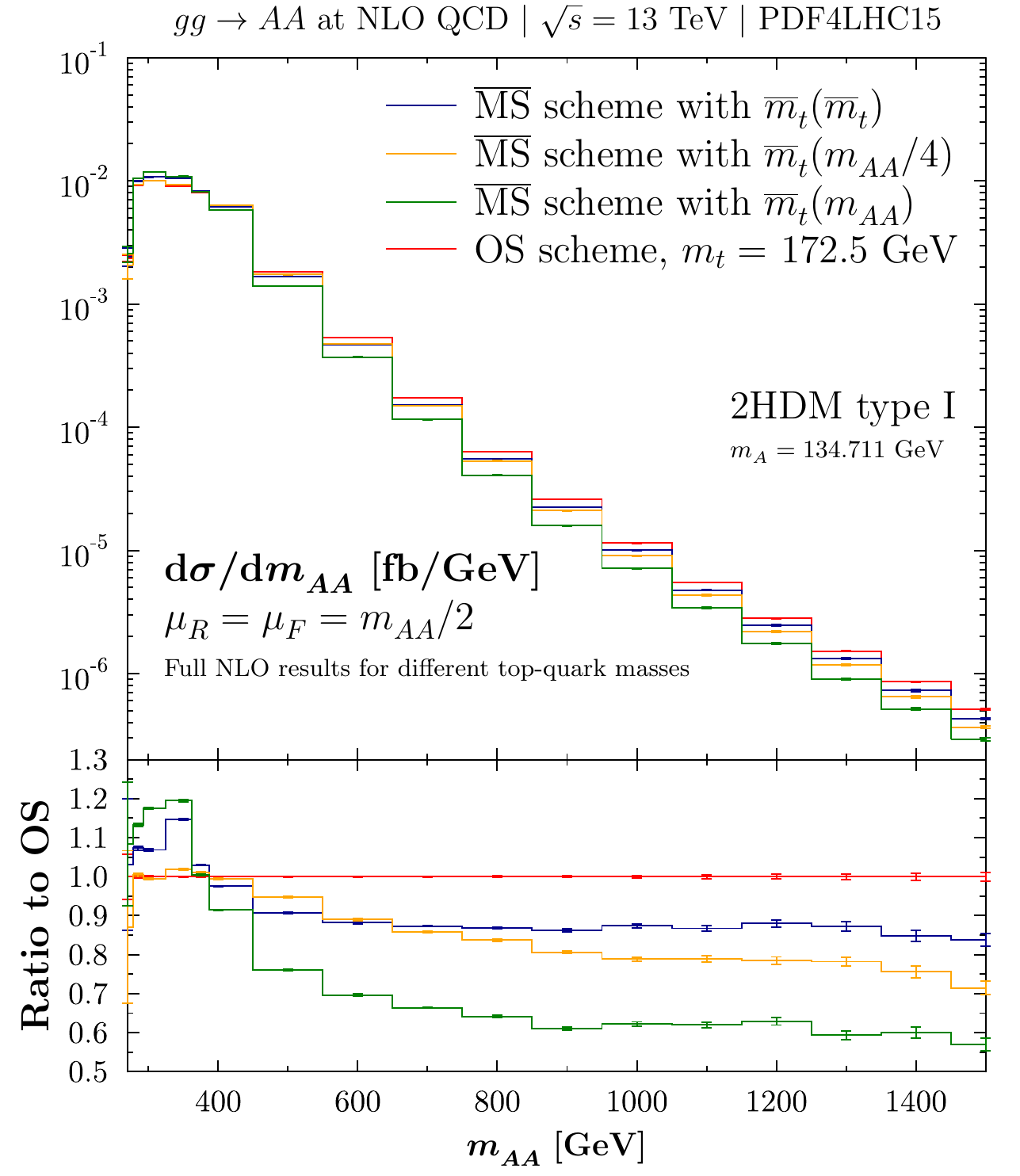}
  \caption[]{Higgs-pair invariant mass distribution at the 13 TeV LHC
    with different scale and scheme for the top-quark mass, in the
    2HDM type I. Left: CP-even $hH$ production. Right: CP-odd $AA$
    production. The lower panels display the ratio to the default
    OS prediction.}
  \label{fig:distribmt13}
\end{figure*}

The calculation of the NLO QCD corrections has been performed in two
different schemes for the renormalisation of the top-quark mass. Our
central predictions use the on-shell (OS) scheme with a mass
$m_t=172.5~\mathrm{GeV}$ both in the Yukawa couplings and in the loop
propagators. The $\overline{\text{MS}}$ scheme can instead
be used, with an appropriate choice of the top-quark mass
counterterm. On top of this scheme choice, there is also a scale choice
for the renormalisation of the top-quark mass,
$\overline{m}(\mu_t)$. To obtain the top-quark scale and scheme
uncertainties, we have compared three $\overline{\text{MS}}$ predictions to
our central OS prediction, for $\mu_t = Q/4$, $Q$, and $\mu_t$ at the
$\overline{\rm MS}$ top mass itself, $\overline{m}_t(\overline{m}_t) =
163.02$ GeV for our choice of the OS top-quark mass value, obtained with an
N$^3$LO evolution and conversion of the pole into the $\overline{\rm
  MS}$ mass $\overline{m}_t(\overline{m}_t)$. The minimal and maximal
cross sections against the central OS prediction are used to calculate
the scale and scheme uncertainties. This procedure has
already been used for SM predictions and this gives rise to significant
uncertainties that are comparable or even larger than the usual
factorisation and renormalisation scale
uncertainties~\cite{Baglio:2018lrj,Baglio:2020ini,Baglio:2020wgt}. 

\begin{figure*}[t!]
  \centering
  \includegraphics[scale=0.63]{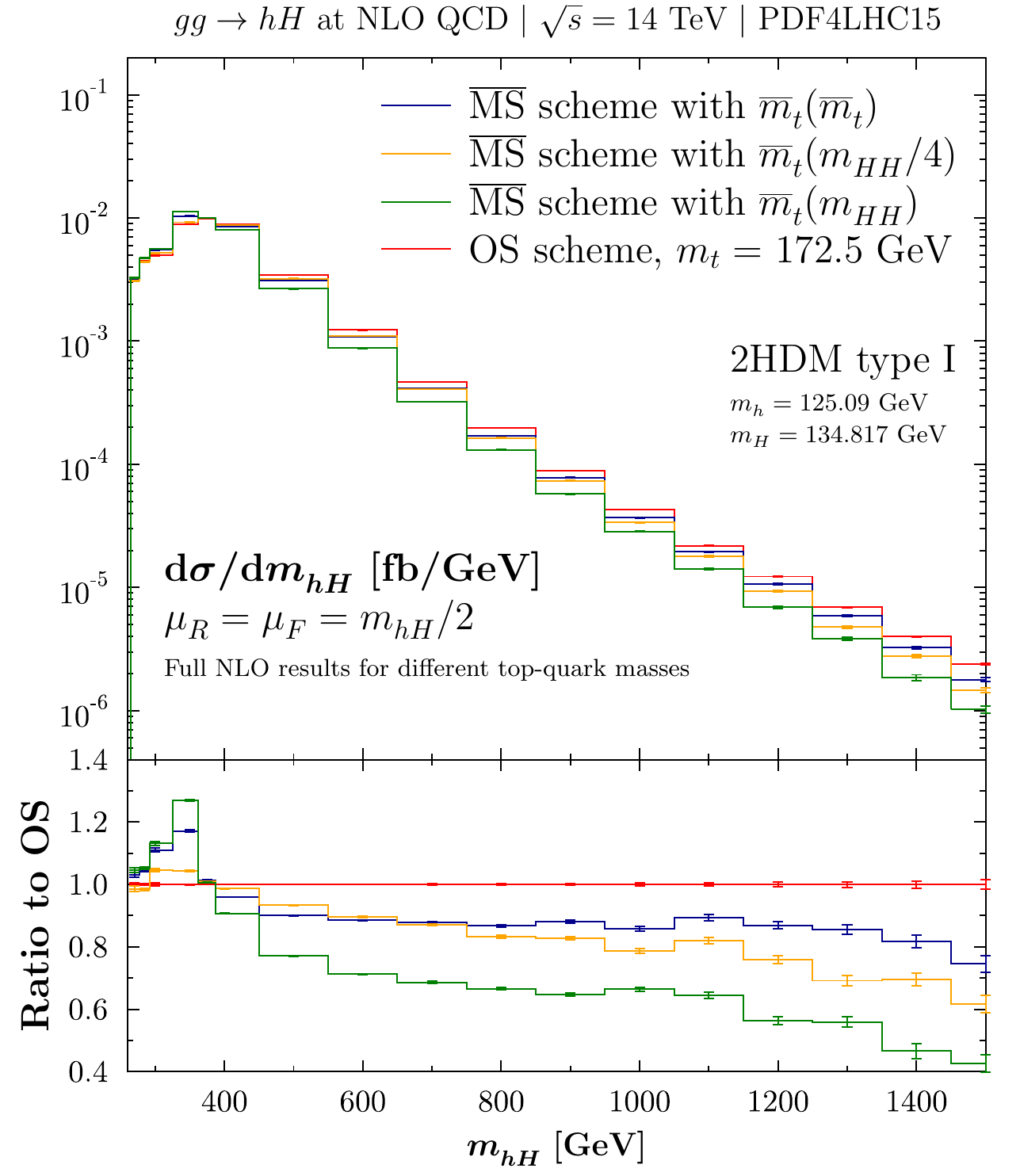}
  \hspace{3mm}
  \includegraphics[scale=0.63]{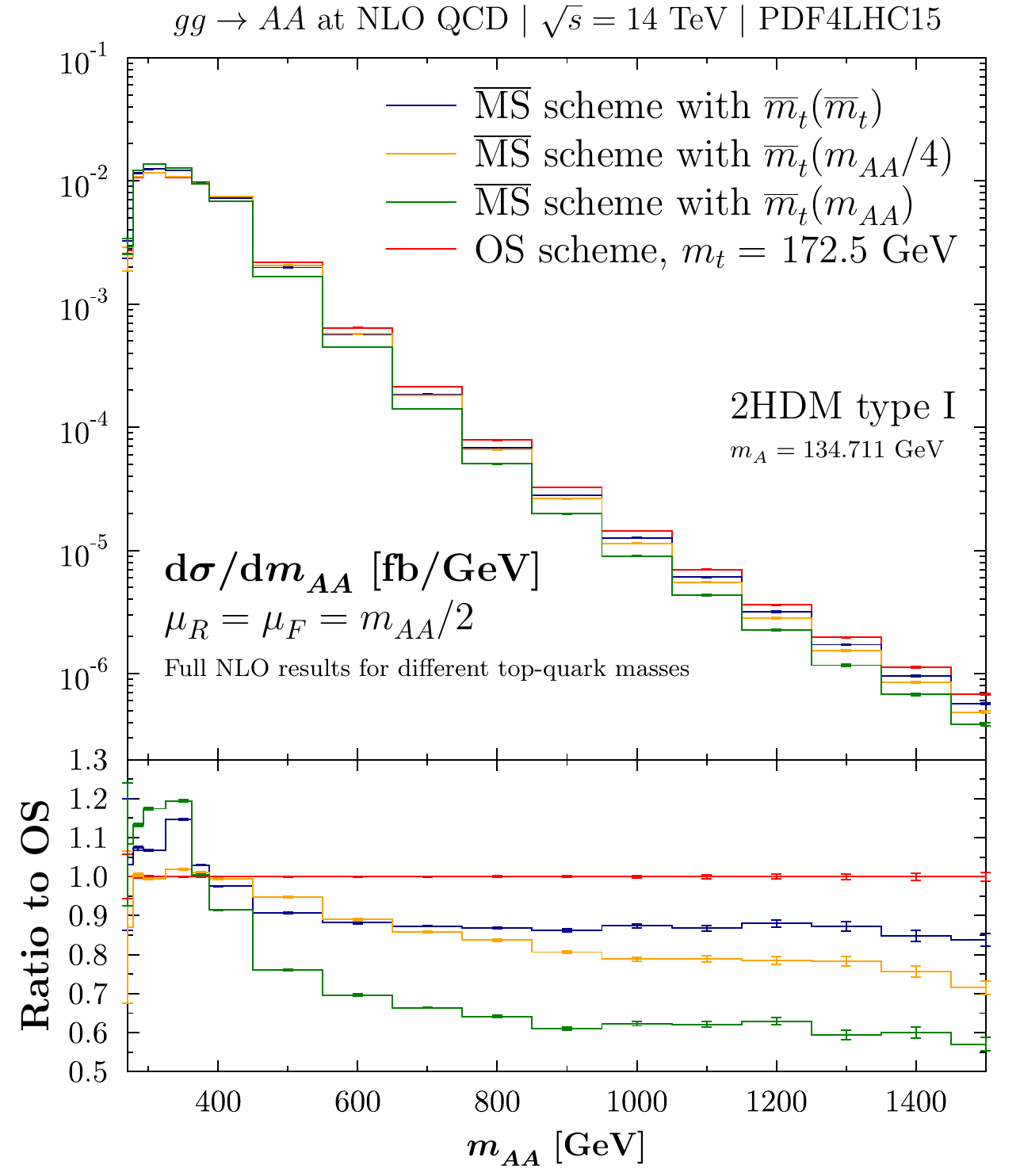}
  \caption[]{Same as in Fig.~\protect\ref{fig:distribmt13} but for $\sqrt{s}=14~\mathrm{TeV}$.}
  \label{fig:distribmt14}
\end{figure*}

\begin{figure*}[t!]
  \centering
  \includegraphics[scale=0.63]{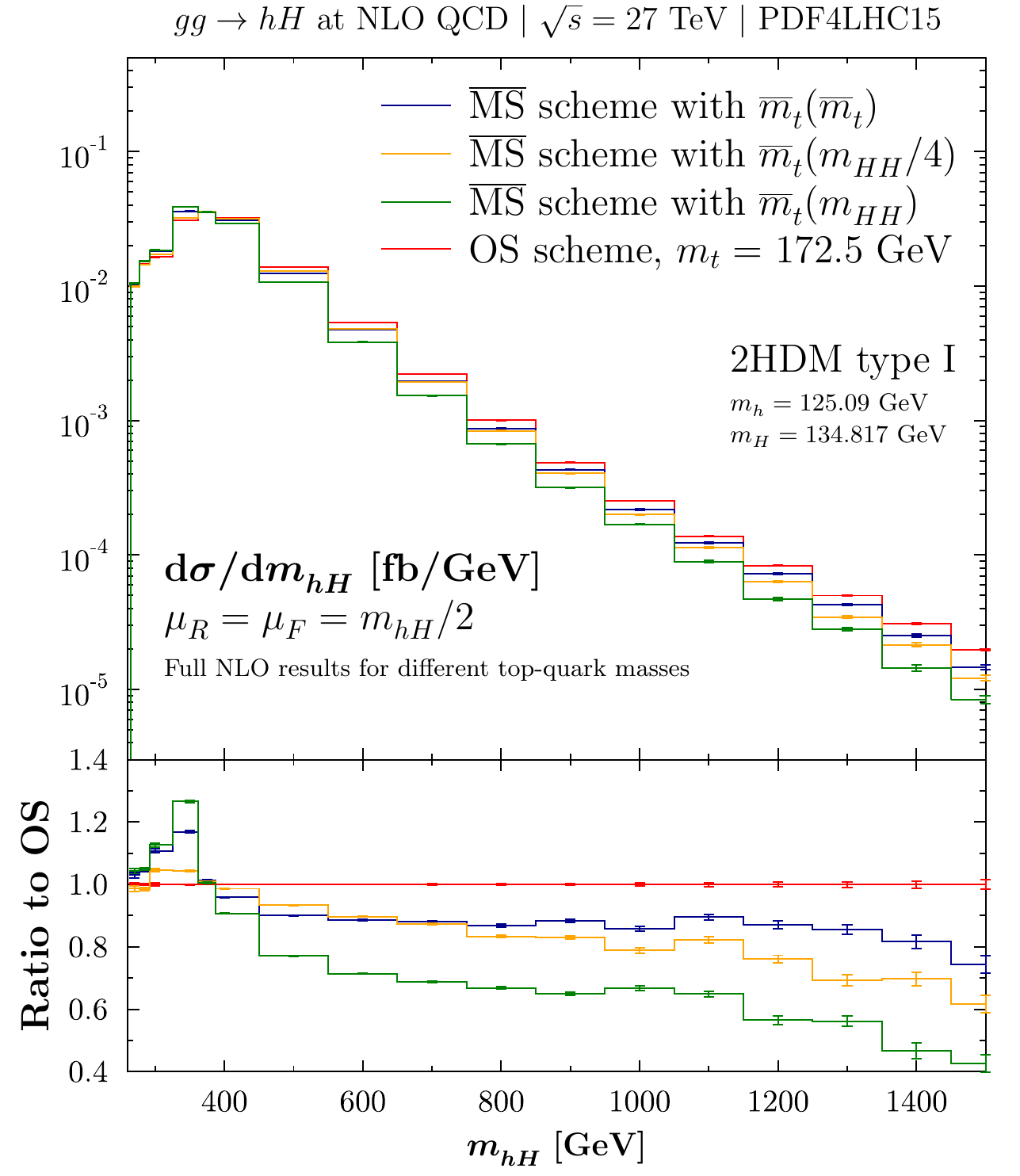}
  \hspace{3mm}
  \includegraphics[scale=0.63]{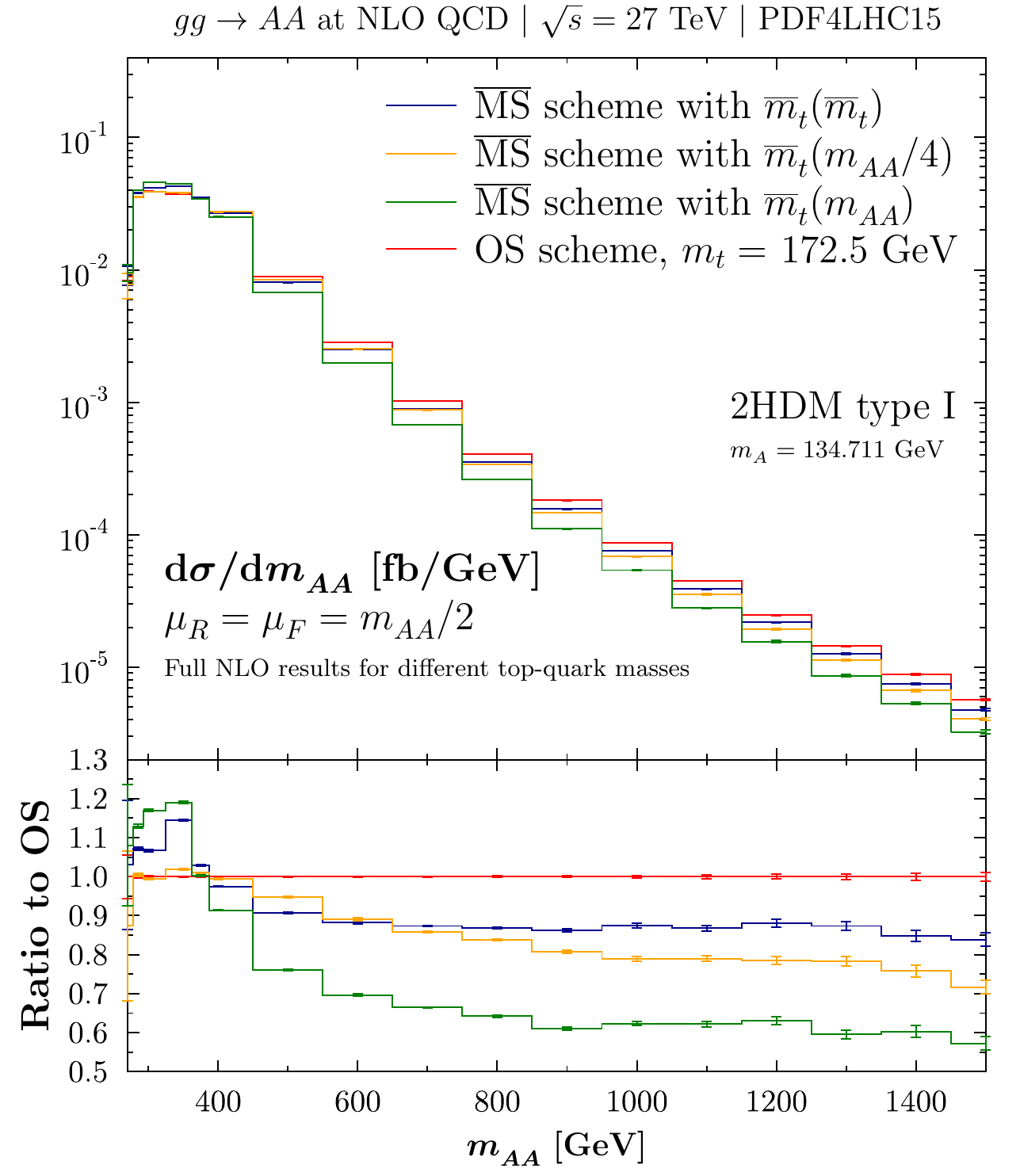}
  \caption[]{Same as in Fig.~\protect\ref{fig:distribmt13} but for $\sqrt{s}=27~\mathrm{TeV}$.}
  \label{fig:distribmt27}
\end{figure*}

\begin{figure*}[t!]
  \centering
  \includegraphics[scale=0.63]{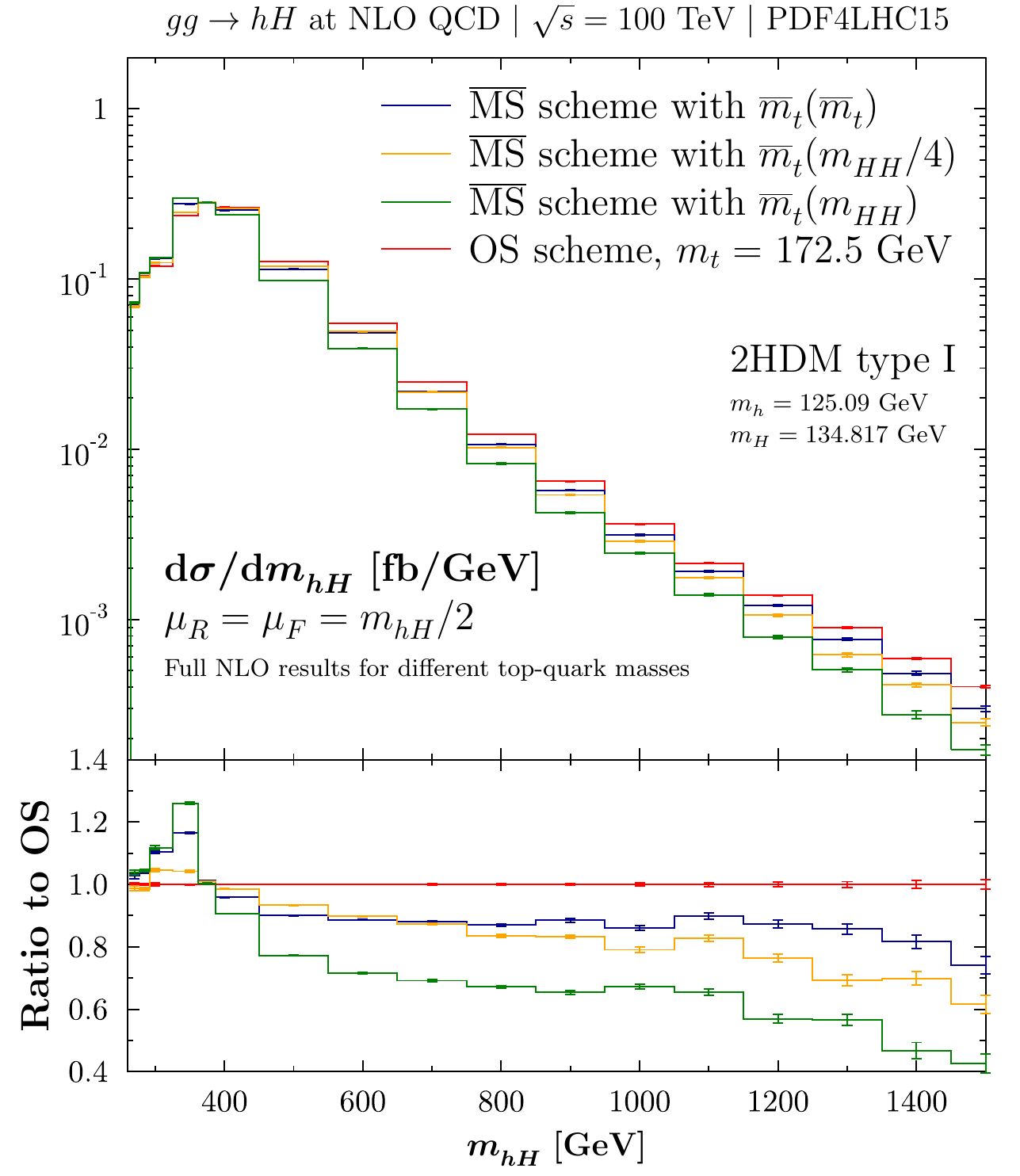}
  \hspace{3mm}
  \includegraphics[scale=0.63]{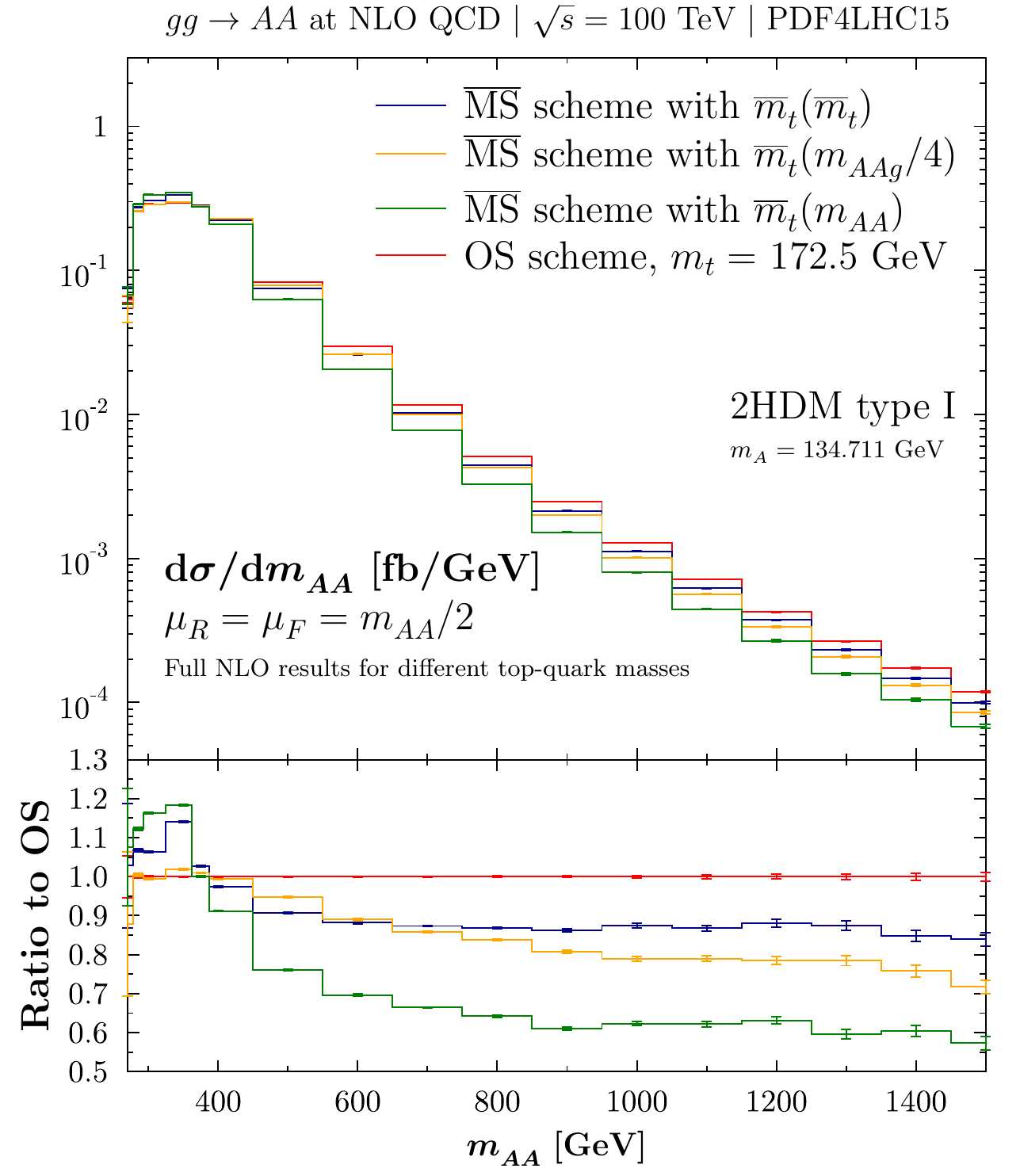}
  \caption[]{Same as in Fig.~\protect\ref{fig:distribmt13} but for $\sqrt{s}=100~\mathrm{TeV}$.}
  \label{fig:distribmt100}
\end{figure*}

We compare the five predictions (the OS predictions and the
three
$\overline{\text{MS}}$ predictions) in Fig.~\ref{fig:distribmt13} at
the 13 TeV LHC, in Fig.~\ref{fig:distribmt14} at
the 14 TeV LHC, in Fig.~\ref{fig:distribmt27} at
the 27 TeV HE-LHC, and in Fig.~\ref{fig:distribmt100} at the 100 TeV
FCC. The red lines display the OS full NLO QCD Higgs-pair invariant
mass distributions, the blue lines the
$\overline{\text{MS}}$ full NLO QCD predictions with
$\overline{m}_t(\overline{m}_t)$, the yellow lines the
$\overline{\text{MS}}$ full NLO QCD predictions with
$\overline{m}_t(Q/4)$, and the green lines exhibit the
$\overline{\text{MS}}$ full NLO QCD predictions with
$\overline{m}_t(Q)$. For $Q$ values above $Q=400$~GeV, the
$\overline{\text{MS}}$ prediction with $\mu_t=Q$
always leads to the
smallest distribution while the maximum at large $Q$ values is given
by the OS prediction.
The lower panels in each figures display the
ratios of the various predictions to our central OS prediction. As
in the SM case, we see large deviations at large $Q$ values,
$\simeq -50\%$ at $Q=1500$~GeV for all c.m. energies.
 We have obtained the following
uncertainties at 13 TeV for selected $Q$ values in $hH$ production
using {\tt PDF4LHC15} parton densities,
\begin{align}
  \frac{d\sigma(gg\to hH)}{dQ}\Big|_{Q=300~{\rm GeV}} & = 0.004278(2)^{+13\%}_{-0\%}\,
                                                        \mathrm{fb/GeV},\nonumber\\
  \frac{d\sigma(gg\to hH)}{dQ}\Big|_{Q=400~{\rm GeV}} & = 0.007522(5)^{+0\%}_{-9\%}\,
                                                        \mathrm{fb/GeV},\nonumber\\
  \frac{d\sigma(gg\to hH)}{dQ}\Big|_{Q=600~{\rm GeV}} & = 0.0010217(9)^{+0\%}_{-29\%}\,
                                                        \mathrm{fb/GeV},\nonumber\\
  \frac{d\sigma(gg\to hH)}{dQ}\Big|_{Q=1200~{\rm GeV}} & = 0.00000956(6)^{+0\%}_{-44\%}\,
                                                         \mathrm{fb/GeV},
\end{align}
and the following uncertainties in $AA$ production,
\begin{align}
  \frac{d\sigma(gg\to AA)}{dQ}\Big|_{Q=300~{\rm GeV}} & = 0.01005(2)^{+17\%}_{-1\%}\,
                                                        \mathrm{fb/GeV},\nonumber\\
  \frac{d\sigma(gg\to AA)}{dQ}\Big|_{Q=400~{\rm GeV}} & = 0.006346(6)^{+0\%}_{-9\%}\,
                                                        \mathrm{fb/GeV},\nonumber\\
  \frac{d\sigma(gg\to AA)}{dQ}\Big|_{Q=600~{\rm GeV}} & = 0.0005328(7)^{+0\%}_{-30\%}\,
                                                        \mathrm{fb/GeV},\nonumber\\
  \frac{d\sigma(gg\to AA)}{dQ}\Big|_{Q=1200~{\rm GeV}} & = 0.00000280(2)^{+0\%}_{-37\%}\,
                                                         \mathrm{fb/GeV}.
\end{align}
As already seen in the SM case, the top-quark scale and scheme
uncertainties turn out to be significant, as large or even larger than
the factorisation and renormalisation scale uncertainties. For
$Q>400$~GeV, the maximum cross section is
always the OS prediction.


From the differential distributions, we can obtain the top-quark 
scale and scheme uncertainties on the total cross section. We adopt
the envelope for each Q-bin individually to build up two maximal and
minimal differential distributions and we integrate these
distributions over $Q$ using fits of the various distributions which
are then numerically integrated. We have arrived at the following top-quark
scale and scheme uncertainties for the CP-even $hH$ total cross
section,
\begin{align}
  13~\mathrm{TeV}: \quad
  \sigma_{gg\to hH}^{} = 1.592(1)^{+6\%}_{-11\%}\,\mathrm{fb}, \nonumber\\
  14~\mathrm{TeV}: \quad
  \sigma_{gg\to hH}^{} = 1.876(1)^{+6\%}_{-11\%}\,\mathrm{fb}, \nonumber\\
  27~\mathrm{TeV}: \quad
  \sigma_{gg\to hH}^{} = 7.036(4)^{+5\%}_{-12\%}\,\mathrm{fb}, \nonumber\\
  100~\mathrm{TeV}: \quad
  \sigma_{gg\to hH}^{} = 60.49(4)^{+4\%}_{-14\%}\,\mathrm{fb},
  \label{eq:hHtoperrors}
\end{align}
and we have obtained the following results for the CP-odd $AA$ total cross
section,
\begin{align}
  13~\mathrm{TeV}: \quad
  \sigma_{gg\to AA}^{} = 1.643(1)^{+9\%}_{-7\%}\,\mathrm{fb}, \nonumber\\
  14~\mathrm{TeV}: \quad
  \sigma_{gg\to AA}^{} = 1.927(1)^{+9\%}_{-8\%}\,\mathrm{fb}, \nonumber\\
  27~\mathrm{TeV}: \quad
  \sigma_{gg\to AA}^{} = 7.012(4)^{+8\%}_{-8\%}\,\mathrm{fb}, \nonumber\\
  100~\mathrm{TeV}: \quad
  \sigma_{gg\to AA}^{} = 58.12(3)^{+7\%}_{-9\%}\,\mathrm{fb}.
  \label{eq:AAtoperrors}
\end{align}
The scale and scheme uncertainties are sizeable and should be included
in an uncertainty analysis of the 2HDM Higgs-pair production cross
sections according to the procedure of Ref.~\cite{Baglio:2020wgt}.

\section{Conclusions}
\label{sec:conclusion}

In this work, we have calculated the full NLO QCD corrections to
mixed scalar
and pure pseudoscalar Higgs-boson pair production
via gluon fusion $gg\to 
hH, AA$ within the 2HDM type I, working in our benchmark
scenario that is not excluded at the LHC.
  We have integrated the
two-loop box diagrams numerically by performing end-point and infrared
subtractions of the contributing Feynman integrals. A numerical
stabilisation across the virtual thresholds has been achieved by
integration by parts of the integrand to reduce the power of the
problematic denominators of the Feynman integrals. The results of the
triangle diagrams, involving s-channel scalar Higgs propagators and
the corresponding trilinear Higgs couplings, have been adopted from the
single-Higgs case. The one-particle reducible contributions emerging from
either two single scalar or pseudoscalar Higgs couplings to gluons can
be derived from the known results for $h,H,A\to Z\gamma$ with
appropriate replacements of the contributing couplings and
masses. After renormalising the top mass and the strong coupling, we
have subtracted the (Born-improved) HTL to obtain the pure virtual NLO
top-mass effects. The real corrections have
been computed by generating the full one-loop matrix elements with
automatic tools. These have then been connected to suitable subtraction
matrix elements in the HTL for the radiation part, but keeping the full
LO top-mass dependence. This could be achieved by suitably projected
4-momenta inside the LO sub-matrix elements. This yields the pure NLO
top-mass effects of the real corrections.

Adding both subtracted virtual and real corrections, we obtain the
full NLO QCD top-mass effects that have then been added to the
(Born-improved) HTL results of Ref.~\cite{Dawson:1998py} by using the
code {\tt Hpair}. Very similar to the corresponding SM calculation of
Refs.~\cite{Borowka:2016ehy, Borowka:2016ypz, Baglio:2018lrj,
  Baglio:2020ini, Baglio:2020wgt}, we find NLO top-mass effects of
about 15--25\% (depending on the collider energy) for the total cross
sections if the top mass is defined as the top pole mass. For the
invariant Higgs-pair mass distribution, the NLO top-mass effects can
reach a level 30--40\% for large invariant mass
values. The larger the hadronic collider energy, the larger NLO top-mass
effects emerge. The renormalisation and factorisation scale dependence
induces uncertainties at the level of 10--15\% for scalar Higgs pairs
and 12--17\% for pseudoscalar Higgs pairs at NLO, i.e.~similar to the SM
case. We have studied the additional theoretical uncertainties
originating from the scale and scheme choice of the virtual top mass and
obtained additional uncertainties of about 5--15\% for scalar
and about 10\% for pseudoscalar Higgs-pair production that are significant
and should be included in future Higgs-pair analyses. These
uncertainties are larger for distributions at large invariant Higgs-pair
masses.

\bigskip

\begin{acknowledgements}
The work of S.G. and M.M. is supported by the DFG Collaborative Research
Center TRR257 ``Particle Physics Phenomenology after the Higgs
Discovery''. F.C. acknowledges financial support by the Generalitat
Valenciana, Spanish Government, and ERDF funds from the European
Commission (Grants RYC-2014-16061, SEJI-2017/2017/019, PID2020-114473GB-100 and PID2020-113334GB-100). The work of 
J.R. is supported by the Italian Ministry of Research (MUR) under grant PRIN 20172LNEEZ.
We acknowledge support by the state of
Baden-W\"urttemberg through bwHPC and the German Research Foundation
(DFG) through Grant No. INST 39/963-1 FUGG (bwForCluster NEMO).
\end{acknowledgements}

{\small
\bibliographystyle{apsrev4-1}
\bibliography{hh_bsm_letter}
}

\end{document}